\documentclass[final,1p,times,11pt]{elsarticle}

\newcommand{\GeV}{\ \text{GeV}}

\newcommand{\km}{\ \text{km}}
\newcommand{\TeV}{\ \text{TeV}}
\newcommand{\kpc}{\ \text{kpc}}

\newcommand{\s}{\ \text{s}}
\newcommand{\cm}{\ \text{cm}}

\newcommand{\be}{\begin{equation}}
\newcommand{\ee}{\end{equation}}
\newcommand{\ds}{{\sf DarkSUSY}}

\biboptions{sort&compress}

\usepackage{amssymb}
\usepackage{amsmath}
\usepackage{url}
\usepackage{booktabs}

\biboptions{sort&compress}
\usepackage{mciteplus}

\journal{Physics of the Dark Universe}

\begin{document}

\begin{frontmatter}

\title{Gamma Ray Signals from Dark Matter:\\ Concepts, Status and Prospects}

\author[HH]{Torsten Bringmann}
\ead{torsten.bringmann@desy.de}

\author[MPPMU]{Christoph Weniger}
\ead{weniger@mpp.mpg.de}

\address[HH]{II.~Institute for Theoretical Physics, University of Hamburg,
Luruper Chaussee 149, DE-22761 Hamburg, Germany}
\address[MPPMU]{Max-Planck-Institut f\"ur Physik, F\"ohringer Ring 6, 80805
Munich, Germany}

\begin{abstract}
  Weakly interacting massive particles (WIMPs) remain a prime candidate for
  the cosmological dark matter (DM), even in the absence of current collider
  signals that would unambiguously point to  new physics below the TeV scale.
  The self-annihilation of these particles in astronomical targets may leave
  observable imprints in cosmic rays of various kinds. In this review, we
  focus on gamma rays which we argue to play a pronounced role among the
  various possible messengers. We discuss the most promising spectral and
  spatial signatures to look for, give an update on the current state of
  gamma-ray searches for DM and an outlook concerning future prospects. We
  also assess in some detail the implications of a potential signal
  identification for particle DM models as well as for our understanding of
  structure formation.  Special emphasis is put on the possible evidence for a
  130 GeV line-like signal that we recently identified in the data of the Fermi
  gamma-ray space telescope.
\end{abstract}

\end{frontmatter}


\section{Introduction}

Evidence for a sizable non-baryonic and cold dark matter  (DM) component  in
the universe derives from an impressive range of unrelated cosmological
observations \cite{1208.3662}, covering distance scales from tens of kpc to several Gpc and
leaving very little room for alternative explanations. On cosmological scales,
DM  contributes a fraction of $\Omega_\chi = 0.229 \pm 0.015$ to the total
energy density of the universe \cite{Komatsu:2010fb}.  Weakly Interacting
Massive Particles (WIMPs) provide a theoretically particularly appealing class
of candidates for the so far obscure nature of DM 
\cite{Jungman:1995df,Bergstrom:2000pn, Bertone:2004pz},
 with the lightest supersymmetric
neutralino often taken as a useful template for such a WIMP.  It is often
argued that the thermal production of WIMPs in the early universe generically
leads to a relic density that coincides with the observed order of magnitude
of $\Omega_\chi$, though this rests on the assumption of a standard
cosmological expansion history and  there exist  well-motivated particle
physics scenarios that predict alternative production mechanisms for WIMP DM
\cite{Feng:2010gw}.  While the LHC non-observation of new particles below the TeV
scale (apart, possibly, from the Higgs boson) has already prompted  doubts 
whether the WIMP DM scenario is still our best bet \cite{Bertone:2010at},  
it must be stressed that electroweak low-energy observables ($g-2$ in particular) 
do  favor new physics contributions not too far above 100\,GeV \cite{Czarnecki:2001pv}.
While this tension starts to considerably disfavor very constrained models of,
e.g., supersymmetry \cite{Bechtle:2012zk}, it may simply be an indication that
the new physics sector which the WIMP belongs to appears at a much smaller
mass scale than any new colored sector.

Attempts to identify WIMP DM can be classified into collider searches for
missing transverse energy,  direct searches for the recoil  of WIMPs off the
nuclei of terrestrial  detectors and indirect  methods that aim at spotting
the products of WIMP self-annihilation. Among possible messengers for such
indirect searches, \emph{gamma rays}
play a pronounced role as they  propagate essentially unperturbed through the
galaxy and therefore directly point to their sources, leading to
distinctive \emph{spatial signatures}; an even more important aspect, as we
will see, is the appearance of pronounced \emph{spectral signatures}.  This
prime role of gamma rays provides our motivation for an updated and dedicated
review on these messengers, which we hope will prove useful and complementary
to existing general reviews on indirect DM searches 
\cite{Cirelli:2010xx,Profumo:2010ya,Lavalle:2012ef}. 
Indeed, the recent indication for a DM signature in
gamma-ray observations of the Galactic center (GC) \cite{Bringmann:2012vr,
Weniger:2012tx} makes such a review extremely timely, and we therefore
dedicate a considerable part of it to discuss in great detail both the status
of the potential signal and its implications.

Gamma rays can either be observed directly from space or, via the showers of
secondary particles they trigger in the atmosphere, indirectly with
ground-based experiments. The former option necessarily implies rather small
effective areas and  an upper bound on the photon energy that can reliably be
resolved, but allows for a large field of view and the observation of gamma
rays at comparably small energies. Particularly promising instruments for the
latter option are imaging Air Cherenkov Telescopes (IACTs) that detect the
Cherenkov light emitted by the shower particles and use efficient image
reconstruction algorithms to determine the characteristics of the primary
photon. These instruments have a limited field of view and a lower energy
threshold set by the need of discriminating photons from the
background of primary muons and hadronic cosmic rays;
their extremely large effective area and rather small field of view make them
ideal for pointed observations. In Table \ref{tab:telescopes}, we pick  
typical examples for space- and ground-based experiments that are
currently operating or planned for the future and compare some basic telescope
characteristics that are particularly relevant for DM searches.  Experiments
that fall into the same broad categories but are not listed explicitly in the
Table include for example AGILE~\cite{Tavani:2008sp} and VERITAS \cite{Holder:2006gi}, as
well as the future CALET \cite{CALET}  and
DAMPE~\cite{DAMPE, Li:2012qg}. We stress that the numbers in
Tab.~\ref{tab:telescopes} are intended to provide a convenient
order-of-magnitude comparison of instrumental characteristics; they should
\emph{not} be used as the basis of detailed sensitivity estimates (see, however, 
the stated references).

\begin{table}[t!]
  \scriptsize
  \centering
  \begin{tabular}{llcccccc}
    \toprule


    & Time of   & $E$-range & $A_\text{eff}$ & \!\!\!\!\!\!Sens.\!\!\! & $\Delta E/E$ & F.O.V. & $\Delta \theta$  \\
    & operation & [GeV]        & [m$^2$]        &  \!\!\!\!\!\! [10$^8$m$^{2}$s]$^{\text{-}1}$\!\!\!     &   [\%]      & [sr]   & [$^\circ$]  \\
    \midrule
    Fermi-LAT   & 2008--$2018^*$  & 0.2--300 & 0.8 & 200 & 11 & 2.4  & 0.2         \\
    AMS-02/Ecal & 2011--$2021^*$  & 10--1000    & 0.2  & 1000  & 3  & 0.4  & 1.0         \\
    AMS-02/Trk  & 2011--$2021^*$  & 1--300      & 0.06 & 1000  & 15 & 1.5  & 0.02        \\
    GAMMA-400   & $2018^*$--\dots & 0.1--3000   & 0.4  & 100 & 1  & 1.2  & 0.02 (0.006)       \\
    \midrule
    MAGIC   & 2009--\dots    & $\gtrsim50$ & $2\!\cdot\!10^4(7\!\cdot\!10^4)$
    & $10(0.2)$    & 20(16)  & 0.003      & 0.17(0.08)  \\
    HESS-II & 2012--\dots    & $\gtrsim30$ & $4\!\cdot\!10^3(10^5$)       &
    4(0.1) & 15(15) & 0.003     & 0.13(0.07) \\
    CTA     & $2018^*$--\dots & $\gtrsim20$ & $5\!\cdot\!10^4(10^6)$       &
    1(0.02) & 20(10)   & $>0.006$ & 0.1(0.06)  \\
    \bottomrule
    &&&&&&& \hfill $^*$ planned
  \end{tabular}
  \caption{\label{tab:telescopes}  Rough comparison of basic telescope
  characteristics relevant for indirect DM searches with gamma rays, for a selection of 
  typical space- and ground-based experiments that are currently operating, shortly upcoming
  or planned for the future. The quoted sensitivity is for point sources at the $5\sigma$ level, after
  1yr  (50 hrs) of space- (ground-) based observations and assuming typical backgrounds. 
Where applicable, numbers refer to photon
  energies at or above $E \simeq 100\GeV$ ($1\TeV$). The angular resolution
  $\Delta\theta$ denotes the $68\%$ containment radius.
  More details in 
    Refs.~\cite{Fermi:performancePASS7} (Fermi-LAT),
  \cite{Jacholkowska:2005nz,Battiston:1999yb,AMS:web} 
  (AMS-02), 
  \cite{Galper:2012fp,Topchiev,Galper:2012ji}
  (GAMMA-400), \cite{Aleksic:2011bx} (MAGIC), \cite{Hess2:Hd2012}
  (HESS-II) and \cite{CTA} (CTA).}
\end{table}

The expected DM-induced gamma-ray flux from a direction $\psi$, averaged over
the opening angle $\Delta\psi$ of the detector, is given by
\be
\label{flux}
  \frac{d\Phi_{\gamma}}{dE_\gamma} (E_\gamma,\psi) = 
  \frac{1}{8\pi} {\int_{\Delta\psi}\frac{d\Omega}{\Delta\psi}\int_\mathrm{l.o.s}
  d\ell(\psi) \rho_\chi^2(\mathbf{r})} \times 
  \left({\frac{\langle\sigma v\rangle_\mathrm{ann}}{m_{\chi}^2} \sum_f
  B_f\frac{dN_\gamma^{f}}{dE_\gamma}}\right) \,,
\ee
where the integration is performed along the line of sight (l.o.s.),
$\langle\sigma v\rangle_\mathrm{ann}$ is the average velocity-weighted
annihilation cross section, $m_\chi$ the mass of the DM particle (for which we
assume $\chi=\bar\chi$), $\rho_\chi$ the DM density, $B_f$ the branching ratio
into channel $f$ and  $N_\gamma^{f}$ the number of photons per annihilation.
An often quoted reference value for $\langle\sigma v\rangle_\mathrm{ann}$
is the so-called `thermal cross section' of $\langle
\sigma v\rangle\sim3\cdot10^{-26}{\rm cm}^3{\rm s}^{-1}$,  which is the
annihilation rate expected for thermally produced WIMPs in the most simple
case (i.e.~$s$-wave annihilation without resonances or co-annihilations
\cite{Griest:1990kh}).
The right part (in parentheses) of Eq.~(\ref{flux})  contains all the
particle physics input and, for the typically very small DM velocities, is
usually sufficiently independent of $v(\mathbf{r})$ that it can be pulled
outside the integrals
(note, however, that this is \emph{not} true for strongly velocity-dependent cross-sections like in 
the case of Sommerfeld enhancement  
\cite{sommerfeld,Hisano:2003ec,Hisano:2004ds,ArkaniHamed:2008qn},
 resonances or $p$-wave annihilation). 
It contains the full \emph{spectral} information that
we will discuss in some detail in Section \ref{sec:spec}. The remaining part,
sometimes referred to as the astrophysical factor (or '$J$-value', with $J\equiv\int d\Omega\int\!\!d\ell\,\rho_\chi^2$), contains in that case the full
information about the \emph{spatial} distribution of the signal and will be
discussed  in Section \ref{sec:spatial}.

We  continue by reviewing in Section \ref{sec:status}  gamma-ray limits on DM
annihilation as well as  the current status of claimed DM signals. The
potentially enormous  implications of a signal identification  for our understanding
of both the underlying particle model and structure formation are then
outlined in Section \ref{sec:implications},
with a focus on  the
intriguing 130\,GeV feature in the direction of the GC.
We discuss future prospects for  the detection of DM with gamma rays in
Section \ref{sec:prospects} and conclude in Section \ref{sec:conc}. For most
of this review, we will assume that DM consists of WIMPs; many aspects,
however, can be applied -- or generalized in a straight-forward way -- to other
cases as well, most notably decaying DM
\cite{Buchmuller:2007ui,Cirelli:2012ut}
 (for which one simply has to replace $\frac12\langle\sigma v\rangle\rho_\chi^2\rightarrow m_\chi\Gamma\rho_\chi$ 
in Eq.~(\ref{flux}), where $\Gamma$ is the decay rate).  Where applicable, we will comment
on this on the way.

\section{Spectral signatures}
\label{sec:spec}

At tree level, DM particles annihilate into pairs of quarks, leptons, Higgs
and weak gauge bosons. The hadronization and further decay of these primary
annihilation products  leads to the appearance of \emph{secondary photons},
mainly through $\pi^0\rightarrow\gamma\gamma$, and the resulting  gamma-ray
spectrum $dN_\gamma^f/dE_\gamma$ can be obtained from event generators like
{\sf Pythia} \cite{Sjostrand:2006za}. Codes like \ds\ 
\cite{ds1,ds2}
 provide
user-friendly numerical interpolations of these spectra, based on a large
number of {\sf Pythia} runs, but there also exist several analytic
parameterizations in the literature 
\cite{Fornengo:2004kj, Cembranos:2010dm,Cirelli:2010xx}.
Secondary photons show a featureless spectrum with a rather soft cutoff at the
kinematical limit $E_\gamma=m_\chi$ and are \emph{universal} in the sense that
$dN_\gamma^f/dx$ (with $x\equiv E_\gamma/m_\chi$) takes a very similar form
for almost all channels $f$ and only very weakly depends on $m_\chi$. A
convincing claim of DM detection based exclusively on this signal, which would
show up as a broad bump-like excess over the often rather poorly understood
astrophysical background, appears generically rather challenging. 

For this reason, it is often much better warranted to focus on the pronounced
spectral features that are additionally expected in many DM models -- not only
because they greatly help to discriminate signals from backgrounds, and hence 
effectively increase the sensitivity of gamma-ray telescopes to DM signals 
\cite{Bringmann:2011ye}, but also
because a detection may reveal a lot about the underlying model for the
particle nature of the DM. Before discussing the various types of spectral
features in  more detail, let us briefly mention further contributions to the
total photon yield that do not give rise to such nice spectral features but
may still visibly change the spectrum (in particular at $E_\gamma\ll m_\chi$).
In models with large branching fractions into $e^+e^-$ pairs, e.g.,
\emph{inverse Compton} scattering of those $e^\pm$ on starlight and the cosmic
microwave background leads to additional gamma rays 
\cite{Baltz:2004bb,Regis:2008ij,Zhang:2008tb}.
 Another source
of additional low-energy photons are electroweak 
\cite{Kachelriess:2007aj,Bell:2008ey,Kachelriess:2009zy,Ciafaloni:2010ti} 
 and strong
\cite{asano:2011ik} radiative corrections, with an additional gauge boson in
the final state; these contributions can actually be quite sizable for DM
masses much larger than the gauge boson mass, in particular if the tree-level
annihilation into a pair of SM particles is suppressed. 

\newpage

\begin{figure}[t]
  \begin{center}
    \includegraphics[width=0.7\textwidth]{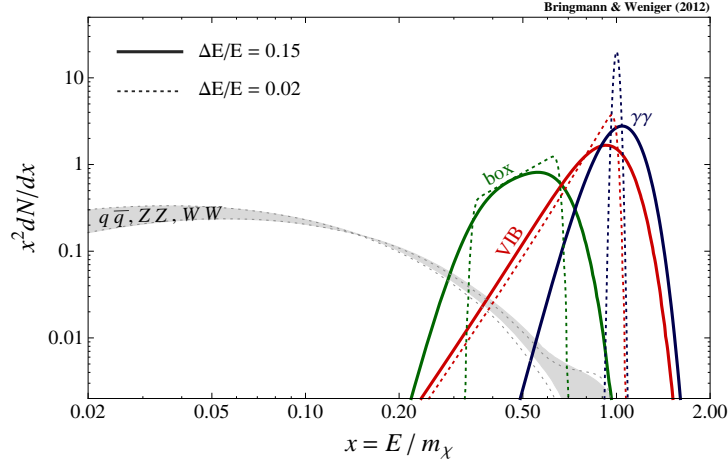}
  \end{center}
  \caption{\label{fig:spec_comp} Various gamma-ray spectra expected from DM
  annihilation, all normalized to $N(x>0.1)=1$.  Spectra from secondary
  particles (gray band) are hardly distinguishable.  Pronounced peaks near the
  kinematical endpoint can have different origins, but detectors with very
  good energy resolutions $\Delta E/E$ may be needed to discriminate amongst them in the
  (typical) situation of limited statistics. See text for more details about
  these spectra.}
\end{figure}

\subsection{Lines}

The direct annihilation of DM pairs into $\gamma X$ -- where $X=\gamma, Z, H$
or some new neutral state -- leads to \emph{monochromatic} gamma rays with
$E_\gamma = m_\chi\left[1-m^2_X/4 m^2_\chi\right]$, providing a striking
signature which is essentially impossible to mimic by astrophysical
contributions \cite{Bergstrom:1997fj}.  Unfortunately, these processes are
loop-suppressed with  $\mathcal{O}(\alpha_{\rm em}^2)$ and thus usually
subdominant, i.e.~not actually visible against the continuous (both
astrophysical and DM induced) background when taking into account realistic
detector  resolutions; however, examples of particularly strong line signals
exist 
\cite{Hisano:2003ec, Hisano:2004ds, Gustafsson:2007pc,Dudas:2009uq,Mambrini:2009ad,Jackson:2009kg,Arina:2009uq}.  
A space-based detector with resolution $\Delta E/E=0.1$
($0.01$) could, e.g., start to discriminate between $\gamma\gamma$ and $\gamma
Z$ lines for DM masses of roughly $m_\chi\lesssim 150\,$GeV ($m_\chi\lesssim
400\,$GeV) if at least one of the lines has a statistical significance of
$\gtrsim5\sigma$~\cite{Bergstrom:2012vd}.  This would, in principle, open the
fascinating possibility of doing `DM spectroscopy' (see also Section
\ref{sec:implications}).

\subsection{Internal bremsstrahlung (IB)}

Whenever DM annihilates into charged particles, additional final state photons
appear at $\mathcal{O}(\alpha_{\rm em})$ that generically dominate the
spectrum at high energies. One may distinguish between \emph{final state
radiation} (FSR) and \emph{virtual internal bremsstrahlung} (VIB) in  a
gauge-invariant way \cite{bringmann:2007nk}, where the latter can very loosely be
associated to photons radiated from charged virtual particles. FSR is
dominated by collinear photons, thus most pronounced for light final state particles,
$m_f\ll m_\chi$, and
produces a model-independent spectrum with a sharp cut-off at
$E_\gamma=m_\chi$ \cite{Beacom:2004pe, birkedal:2005ep}; a typical example for
a spectrum dominated by these contributions is Kaluza-Klein DM
\cite{bergstrom:2004cy}. VIB, on the other hand, dominates if the tree-level
annihilation rate is suppressed (like e.g.~the annihilation of Majorana
particles into light fermions 
\cite{Bergstrom:1989jr,Flores:1989ru})
 and/or
the final state consists of bosons \emph{and} the $t$-channel particle is
almost degenerate with $m_\chi$ \cite{Bergstrom:2005ss}. It generates
pronounced bump-like features at $E_\gamma\lesssim m_\chi$ which closely
resemble a slightly distorted line for energy resolutions $\Delta E/E\gtrsim
0.1$. The exact form of VIB spectra, however, is rather
model-dependent  \cite{bringmann:2007nk} -- which in principle would allow an 
efficient discrimination between DM models for large enough statistics 
(see e.g.~\cite{Perelstein:2010at}).

\subsection{Cascade decays}

Another possibility to produce pronounced spectral features is DM annihilating
into  intermediate neutral states, $\chi\chi\rightarrow\phi\phi$, which then
decay directly ($\phi\rightarrow\gamma\gamma$ \cite{Ibarra:2012dw}) or via FSR
(e.g.~$\phi\rightarrow \ell^+\ell^-\gamma$ \cite{Bergstrom:2008ag}) into
photons. While the latter situation results in a spectrum that resembles the
standard FSR case (with a slightly less sharp cut-off and a potentially
considerably reduced rate in the degenerate case, $m_\phi\sim m_\ell$), the former process induces
a box-shaped spectrum with a width of $\Delta E=\sqrt{m_\chi^2-m_\phi^2}$; for
small mass differences, it is thus indistinguishable from a line.

\bigskip
In Fig.~\ref{fig:spec_comp}, we  compare the various different spectra
discussed above; in order not to overload this figure, however, we do not
include the FSR-dominated spectrum from lepton final states (see e.g.~Ref.~\cite{bergstrom:2004cy}). 
Secondary photon spectra from all possible
quark or weak gauge boson final states are all contained in the rather thin
gray band (we adopted $m_\chi=100\,$GeV, though the result is quite insensitive
to this value). For the VIB spectrum, we assumed Majorana DM annihilation into
light fermions via a scalar $t$-channel particle (`sfermion') almost
degenerate in mass with $\chi$, like encountered in supersymmetry
\cite{bringmann:2007nk}, and for the box we chose $m_\phi=0.95m_\chi$.


\newpage

\section{Spatial Signatures} 
\label{sec:spatial}
The peculiar morphology of annihilation signals, tracing directly the DM density,
offers another convenient handle for discriminating signals from backgrounds.
 The most relevant targets are the GC, dwarf
spheroidal galaxies and galaxy clusters with respective half light radii of
roughly $\theta_{1/2}\lesssim 10^\circ$, $\theta_{1/2} \sim 0.1^\circ$ and
$\theta_{1/2} \gtrsim 0.1^\circ$.
Further important targets are DM  clumps
or the angular power spectrum of the isotropic gamma-ray background (IGRB),
all of which we will discuss in this section.

\subsection{Halo Profiles and the Galactic Center}
\label{sec:GC}
The arguably brightest source of gamma rays from DM annihilation is the center
of our Galaxy. Within a few degrees (say $2^\circ\times2^\circ$) around the GC, WIMPs  would induce a 
gamma-ray flux of about
$\mathcal{O}(10^{-8}) \rm\ ph\ cm^{-2}\ s^{-1}$ at the Earth (at $>1\GeV$,
assuming a thermal annihilation rate into $\bar{b} b$, $m_\chi=100\GeV$ and standard
halo profiles), very well in reach of current instruments.  However, the
line-of-sight to the GC traverses the galactic disc, which harbours numerous
 high-energetic processes ($\pi^0$ production in cosmic-ray interactions,
Bremsstrahlung and inverse Compton emission, bright point sources); the
corresponding gamma-ray fluxes of $\mathcal{O}(10^{-7}$--$10^{-6}) \rm\ ph\ cm^{-2}\
s^{-1}$ thus outshine DM signals often by orders of magnitude.
Furthermore, the uncertainties in the signal and background morphologies make
the identification of a DM signal from the inner Galaxy a challenging task.

A useful general parametrization of DM halos, which encompasses a
large number of commonly used profiles, reads
\begin{align}
  \rho^{\alpha\beta\gamma}_{\chi}(r) = \rho_\odot \left[ \frac{r}{r_\odot}
  \right]^{-\gamma} \left[
  \frac{1+(r_\odot/r_s)^\alpha}{1+(r/r_s)^\alpha}
  \right]^{\frac{\beta-\gamma}{\alpha}}\,,
  \label{eqn:abgProfile}
\end{align}
where $r$ is the distance from the halo center, $r_\odot\simeq8.5\kpc$ 
the position of the Sun and $\rho_\odot\simeq0.4\GeV\cm^{-3}$  the local 
DM density 
\cite{Catena:2009mf,Salucci:2010qr,McMillan:2011wd,Bovy:2012tw}
 (see Ref.~\cite{Pato:2010yq}
for a discussion of systematic uncertainties of this quantity and Ref.~\cite{Garbari:2012ff} for
a recent study that includes the effect of a slightly oblate DM halo and the possible presence
of a dark disc, leading to a normalization which is a factor of $\sim$2-3 larger). 
The parameters $\alpha$, $\beta$ and $\gamma$ determine the halo shape, and
$r_s$ the concentration. The
commonly used Navarro-Frenk-White (NFW)
profile~\cite{Navarro:1995iw}, for example, is obtained for $(\alpha, \beta, \gamma) =
(1,3,1)$ (with $r_s\simeq20\kpc$ in case of the Milky Way); the cored
isothermal profile follows when setting $(\alpha, \beta, \gamma) = (2,2,0)$
and $r_s\simeq3.5\kpc$ (note, however, that matching observational constraints in 
principle results in a rather large range of allowed values for $r_s$  
\cite{Klypin:2001xu,Widrow:2005bt}).
Typically, kinematic observations of line-of-sight velocities do not sufficiently
constrain the DM profile, so one has to rely on (extrapolated) results from
numerical $N$-body simulations of dissipationless structure formation in a 
$\Lambda$CDM cosmology \cite{Diemand:2008in, Springel:2008cc}; 
see also Ref.~\cite{Diemand:2009bm,Kuhlen:2012ft} for a review. 
Most recent results tend to favor a spherical Einasto profile 
\cite{Einasto:1965},
\begin{equation}
  \rho_\chi^{\rm Einasto}(r)= \rho_\odot \exp \left( -\frac{2}{\alpha_E}
  \frac{r^{\alpha_E}-r_\odot^{\alpha_E}}{r_s^{\alpha_E}} \right)\;,
  \label{einasto}
\end{equation}
over the somewhat steeper NFW profile. Both $r_s$ and the halo shape
parameter $\alpha_E$ depend on the total halo mass; in case of
the Milky Way, numerical simulations yield $\alpha_E\simeq0.17$ and
$r_s\simeq20\kpc$~\cite{Diemand:2008in}. Note that for $r\lesssim 0.01r_s$, there are actually numerical indications for profiles with logarithmic slopes of about $\gamma\sim1.2$, i.e.~steeper  than both NFW and Einasto 
\cite{Diemand:2005wv,Stadel:2008pn,Diemand:2008in}.

In the Milky Way, baryonic matter dominates the gravitational potential for roughly
$r\lesssim r_\odot$, which can have a great impact on the DM distribution 
with respect to the expectations from DM-only simulations mentioned above. 
In particular, the cooling and
infall of baryons could -- by a mechanism known as adiabatic contraction -- lead
to a steepening of the inner DM profile
\cite{Blumenthal:1985qy,Gnedin:2004cx,Gustafsson:2006gr}.

However, such a scenario becomes much less likely if, as sometimes 
found in simulations \cite{Governato:2009bg}, feedback from star 
formation and supernovae dominates over cooling and infall processes.
In fact, the presence of baryons could 
have the opposite effect of producing cores rather than cusps, see 
e.g.~the discussion in 
Ref.~\cite{Primack:2009jr,deBlok:2009sp,Kuhlen:2012ft}, 
even for a system with the size of the Milky Way \cite{Governato:2012fa}
(though such a core might also form only in the very center of a contracted profile \cite{Guedes:2011ux}).
 The
adiabatic growth of the central supermassive black hole (SMBH) could also generate a
central spike of DM within the inner $\sim 10\ \rm pc$~\cite{Ullio:2001fb} if
the SMBH seed starts out close to the GC~\cite{Gondolo:1999ef} (SMBH
mergers, on the other hand, would rather \emph{destroy} initial cusps in the profile
\cite{Merritt:2002vj}).
Microlensing and stellar rotation curve observations can only exclude the most
extreme scenarios, yielding upper limits of about $\gamma\lesssim 1.5$ for the 
logarithmic slope of the DM profile near the GC \cite{Iocco:2011jz}.
The resulting difference in the expected DM annihilation flux from the inner $\sim
0.1^\circ$ around the GC, when comparing the most extreme cases of a cored profile 
and a profile as steep as this upper limit, amounts to around five orders of 
magnitude \cite{Fornengo:2004kj}. 
In any case, the annihilation signal from the GC would most likely appear as an extended
source with a peculiar angular
profile
\cite{Zaharijas:2006qb,Dodelson:2007gd}. 
 Due to the large
astrophysical foreground in the very center (e.g.~the bright HESS source
J1745-290~\cite{Aharonian:2006wh}), the optimal region of interest (ROI) for
signal searches extends out to a few degrees and could also lie slightly away from the GC~\cite{Serpico:2008ga,Bringmann:2011ye,Bringmann:2012vr}.

A topic that has received only little attention is the possibility that the point
of highest DM density could be displaced from the GC.  The latter lies at the
dynamical center of our Galaxy, coinciding with the position of Sgr A$^\ast$.
Baryons are directly affected by star formation and supernovae, and they shock
during galaxy mergers whereas DM does not.  It is not unlikely that at least
during some periods in the merger and formation history of our Galaxy
significant displacements existed~\cite{Ullio:2001fb}, like
{e.g.}~observed for the SMBH in M87~\cite{Batcheldor:2010vd}. If such a displacement
remained until today, it could provide a spectacular window into the formation
history of our Galaxy, but would also introduce another unknown into the
search for a DM signal. Actually, state-of-the-art simulations of late-time
spiral galaxies with a significant bar, like the Milky Way, show first hints for such a
displacement~\cite{Guedes:2011ux, kuhlen}.

\subsection{Substructure Enhancement} 

For annihilating DM, the gamma-ray emissivity is proportional to the DM density
squared. Unresolved substructures in the DM distribution, predicted to exist by all 
cold DM $N$-body simulations,  can hence potentially have a huge
impact on the signal strength \cite{Silk:1992bh} -- simply because the 
astrophysical factor in Eq.~(\ref{flux}) effectively averages over $\rho_\chi^2$ 
and one always has $\langle\rho_\chi^2\rangle>\langle\rho_\chi\rangle^2$ for inhomogeneous distributions. This effect is typically quantified in terms of the
so-called boost factor $B$, which can be defined as the ratio of the actual line-of-sight integral to
the one obtained assuming a smooth (e.g.~Einasto) component only. 

Roughly, every decade in subhalo masses down to the cutoff scale contributes
the same to the total gamma-ray flux~\cite{Diemand:2006ik} (with small subhalos 
possibly being slightly more important \cite{Diemand:2009bm}), though the details
depend crucially on the adopted subhalo mass distribution and concentration,
as well as survival probabilities of the smallest clumps
 (all of which have to be extrapolated over many orders of magnitude from
 the results of $N$-body simulations). 
The lower cutoff in the subhalo mass distribution is set by the kinetic
decoupling of WIMPs in the early Universe \cite{Green:2005fa} and strongly
depends on the DM particle properties \cite{Profumo:2006bv, Bringmann:2009vf};
while in principle it could be as large
as the scale of dwarf galaxies \cite{Aarssen:2012fx}, it falls into the  range of roughly
$10^{-11}$--$10^{-3}M_\odot$ for standard MSSM 
neutralinos  \cite{Bringmann:2009vf} . 

Not too much
is known about the precise distribution of substructures, but what one can learn
from $N$-body simulations is that due to merger and tidal stripping in
the Milky Way halo (see e.g.~\cite{Springel:2008cc,Pieri:2009je} and
references therein), fewer substructures are expected in the inner galaxy than
in the outer parts (though the surviving subhalos close to the halo center have larger
concentrations). 
This implies that the expected boost factor for GC observations is of order unity
\cite{Diemand:2008in,Kuhlen:2008aw,Pieri:2009je,Kamionkowski:2010mi}, 
while it may be as large as $\mathcal{O}(1000)$
for galaxy clusters due to the enormous contained hierarchy of masses 
\cite{Jeltema:2008vu, SanchezConde:2011ap, Gao:2011rf, Pinzke:2011ek}. 
Note also that the expected angular dependence of the signal can change in
the presence of large boost factors: in the limit where unresolved substructures completely
dominate the total signal, the flux essentially scales with (the line-of-sight integral over)
$\rho_\chi$ rather than $\rho_\chi^2$;
for the Milky Way, this implies that the halo flux emissivity could be changed for 
$r\gtrsim1$\,kpc \cite{Bergstrom:1998jj,Kuhlen:2008aw,Pieri:2009je}.

\subsection{Point-like Sources}
Many complications associated with the GC are avoided when looking at
point-like targets outside the galactic disk. The corresponding signals
are typically considerably fainter, which is however compensated by the greatly
simplified and much smaller astrophysical background.

The probably most promising source class are nearby dwarf spheroidal 
galaxies.  These faint 
satellites of the Milky Way exhibit the largest known mass-to-light ratios, up to
$\sim1000\ M_\odot/L_\odot$, and do not show signs for gas or recent star
formation. As such, they are not expected to be gamma-ray emitters
\cite{Baltz:1999ra,Tyler:2002ux,Evans:2003sc}.
 A canonical set of
less than ten dwarf spheroidals were subject of numerous recent
studies
\cite{Bringmann:2008kj,Pieri:2008nb,Martinez:2009jh, Scott:2009jn}. 
Since dwarf spheroidals are DM dominated, stellar kinematics can be efficiently 
used to constrain the DM content
\cite{Strigari:2007at, Charbonnier:2011ft,Walker:2011fs}.  
Despite the remaining uncertainties in the shape of the DM
profile, it turns out that the integrated signal fluxes are surprisingly
robust: the  uncertainty is at the level of only $10\%$--$50\%$ \cite{Ackermann:2011wa}
under the \emph{assumption} of an NFW-like profile in the central part,
which is at least naively well motivated (following the idea that such highly DM dominated
systems might follow the expectations from CDM only simulations rather closely) and 
also consistent with observations \cite{Strigari:2010un} 
(but see~\cite{Gilmore:2007fy}). However, baryonic processes may still change an 
initial NFW profile; allowing the inner slope of the density profile to vary between
$0\leq\gamma\leq1$, e.g., introduces uncertainties corresponding to a factor of a few 
in the resulting flux, while very steep profiles with $\gamma\geq1.5$ would increase 
the flux by an order of magnitude 
\cite{Charbonnier:2011ft,Walker:2011fs}.
Significant substructure boosts seem unlikely  
\cite{Charbonnier:2011ft,Walker:2011fs}
though it is has been speculated  that they might enhance the signal
by up to two orders of magnitude in the most optimistic case \cite{Strigari:2006rd}. 
This situation is in sharp contrast to the large uncertainties
related to the signal flux from the GC that we discussed above and makes dwarf
spheroidal galaxies excellent targets to derive robust constraints on the DM
annihilation cross-section. Furthermore, it allows a simultaneous analysis of
multiple dwarf galaxies, which further increases the sensitivity for DM
signals.

Galaxy clusters are the most massive DM dominated virialized objects in the
Universe and  provide excellent targets to search for an annihilation
signal~\cite{Jeltema:2008vu, Colafrancesco:2005ji}. Harboring an enormous 
hierarchy of substructures, they are the astronomical targets that are expected to 
maximize the boost factor \cite{Pinzke:2011ek}.
In optimistic scenarios~\cite{Pinzke:2011ek},
they could therefore outshine a signal from local dwarf spheroidals by a factor of a
few~\cite{Huang:2011xr, Ackermann:2011wa}, which makes them very attractive
as targets for the potential  \emph{detection} of a DM signal 
\cite{Jeltema:2008vu,SanchezConde:2011ap, Gao:2011rf, Pinzke:2011ek}.  
The rather large
involved astrophysical uncertainties connected to both the subhalo distribution and
cosmic-ray induced gamma rays, on the other hand, imply that robust \emph{limits}
derived from cluster observations are usually not competitive. 
 Similar to dwarfs, the sensitivity to cluster signals can be enhanced by a combined
analysis
\cite{Huang:2011xr, Zimmer:2011vy, Nezri:2012tu,Combet:2012tt}. 

Given that they come with
no intrinsic astrophysical backgrounds, clumps (subhalos) of DM in the galactic halo that are not massive enough to
trigger star formation are further important targets for indirect DM
searches~\cite{Lake:1990du,Pieri:2007ir, Pieri:2009je}. If discovered as
unidentified sources with no counterparts at other wavelengths by surveying
instruments like the Fermi-LAT
\cite{Zechlin:2011kk,Zechlin:2012by}, 
detailed follow-up
observations with IACTs could become vital to prove their DM nature by means
of additional angular and spectral information. 


\subsection{Extragalactic Diffuse Signal and Anisotropies} 

Gamma rays from DM annihilation at cosmological distances,
integrated over all redshifts and (sub)halo distributions,  appear largely
isotropic and add up to the astrophysical IGRB 
together with contributions from galactic DM annihilation 
\cite{Bergstrom:2001jj, Ullio:2002pj,Taylor:2002zd,Blanchet:2012vq,Cirelli:2009dv,Abazajian:2010zb}. 
The astrophysical IGRB is
presently not very well understood, but
believed to stem from unresolved sources like blazars, star-forming galaxies
and milli-second pulsars. A WIMP with thermal annihilation cross-section and
$\mathcal{O}(100\GeV)$ masses could contribute between $\sim1\%$ and
$\sim100\%$ to the IGRB, depending on the minimum mass, concentration and 
abundance of
subhalos~\cite{Abdo:2010dk} (though the intrinsic uncertainties related to 
the substructure distribution may well be even larger
\cite{Serpico:2011in}).  In case of annihilation into $\gamma\gamma$, the
signal would be a gamma-ray line broadened by the redshift and provide a
peculiar spectral signature to look for in the IGRB~\cite{Bergstrom:2001jj},
which can be efficiently constrained by observations~\cite{Abdo:2010dk}.

An interesting approach towards signal identification is also to exploit its
angular power spectrum
\cite{Ando:2005hr,Ando:2006mt,Cuoco:2006tr,Taoso:2008qz,SiegalGaskins:2008ge,
SiegalGaskins:2009ux,Ando:2009fp,Ibarra:2009nw,Cuoco:2010jb,Zavala:2011tt,
Ando:2005xg,Ando:2006cr, Fornasa:2009qh,Lee:2008fm,Zavala:2009zr,
Serpico:2011in, Fornasa:2012gu},
 which receives contributions
from subhalos within our own Galaxy as well as from extragalactic (sub)halos. Let us 
mention here in particular that anisotropy measurements might turn out to be a feasible way 
to probe the minimal subhalo mass 
\cite{Ando:2005xg,Ando:2006cr,Fornasa:2009qh,Lee:2008fm},
which would open a completely complementary window into the particle nature of the DM \cite{Bringmann:2009vf}.

\section{Current status}
\label{sec:status}

\subsection{Limits}

The total number of photons above the detector threshold, typically dominated
by secondary photons, is a very convenient and simple measure to constrain
possible exotic contributions to observed gamma-ray fluxes. Limits on the DM
annihilation rate are therefore usually presented in the $\langle \sigma
v\rangle$ vs. $m_\chi$ plane, with the assumption of WIMPs  dominantly
annihilating into $\bar b b$ being an often adopted standard that is useful
for comparison. Such constraints have been derived from the observation of galaxy
clusters 
\cite{Ackermann:2010rg,Dugger:2010ys, Huang:2011xr, Zimmer:2011vy,
Ando:2012vu, Han:2012uw},
external galaxies 
\cite{Vassiliev:2003cd,Gotting:2003dh,Lavalle:2006rs, Wood:2008hx,Fornengo:2004kj},
 globular clusters \cite{Abramowski:2011hh,Feng:2011ab},
Milky Way satellite dwarf galaxies 
\cite{Scott:2009jn,Abdo:2010ex,Aleksic:2011jx,Mazziotta:2012ux, Ackermann:2011wa,GeringerSameth:2011iw, Cholis:2012am, Acciari:2010ab,Essig:2009jx, Essig:2010em,Wagner:2009wp, Grube:2012fv}, 
the GC
\cite{Morselli:2010ty, Crocker:2010gy, Hooper:2011ti,Buckley:2012ws,Abramowski:2011hc,Cholis:2012fb, Hooper:2012sr}
 and halo \cite{Zaharijas:2010ca,Ackermann:2012rg}, or the IGRB
\cite{Abdo:2010dk, Abazajian:2010zb,Calore:2011bt, Ackermann:2012uf}.  

The currently best limits of this kind, for WIMP masses $5\,{\rm
GeV}\lesssim m_\chi\lesssim1\,{\rm TeV}$, derive from observations of nearby
dwarf  galaxies by the Fermi satellite \cite{Ackermann:2011wa}\footnote{
Depending on the assumed profile and small-scale cutoff (as well as subhalo properties), 
the  recently presented constraints from galaxy clusters \cite{Han:2012uw} are nominally 
even tighter. There are also claims that limits from GC \cite{Hooper:2011ti,
Hooper:2012sr} or globular cluster \cite{Feng:2011ab} observations are actually stronger (though seemingly much less robust) 
than the dwarf limits.}; for
$m_\chi\lesssim 25\,$GeV, these limits are actually stronger than the `thermal' rate
of $\langle\sigma v\rangle_{b\bar b}\sim3\cdot10^{-26}{\rm cm}^3{\rm s}^{-1}$.  
At $m_\chi\simeq700\,$GeV, they weaken to $\langle \sigma
v\rangle_{b\bar b}\lesssim 4\cdot10^{-25}{\rm cm}^3{\rm s}^{-1}$ and for even
higher WIMP masses, the currently strongest limits are presented by the HESS
collaboration from observations of the GC region \cite{Abramowski:2011hc}: at
$m_\chi\sim1\,$TeV ($10\,$TeV), those are about a factor of 10  (30) weaker
than the thermal value (see also Ref.~\cite{GeringerSameth:2011iw,
Abazajian:2011ak} for independent studies of these limits). When comparing
limits from different targets, however, one should always keep in mind that
the underlying astrophysical uncertainties that enter as the line-of-sight
integral in Eq.~(\ref{flux}) may be quite different; in particular, as
stressed in Section \ref{sec:spatial}, predictions for integrated signal
fluxes are much more robust for dwarf galaxies than for the GC.

There have also been various searches for \emph{line signals}: in M31 with
HEGRA \cite{Aharonian:2003xh}, at the GC with EGRET~\cite{Pullen:2006sy}, and
with Fermi-LAT GC \cite{Abdo:2010nc, Vertongen:2011mu, Ackermann:2012qk} as
well as dwarf data \cite{GeringerSameth:2012sr} and in galaxy clusters
\cite{Huang:2012yf}. The currently strongest limits presented by the LAT
collaboration follow from Fermi observations of the GC region
\cite{Ackermann:2012qk} and extend from $\langle \sigma
v\rangle_{\gamma\gamma}\lesssim 3\cdot10^{-29}{\rm cm}^3{\rm s}^{-1}$ at
$m_\chi=10\,$GeV to $\langle \sigma v\rangle_{\gamma\gamma}\lesssim
4\cdot10^{-27}{\rm cm}^3{\rm s}^{-1}$ at $m_\chi=200\,$GeV (slightly stronger
limits can be found in independent analyses~\cite{Vertongen:2011mu,
Weniger:2012tx} for masses from 1 to 300 GeV).  
Preliminary results from HESS
exclude lines above $500\GeV$ down to cross-sections of $\langle\sigma
v\rangle_{\gamma\gamma}\sim2\times10^{-27}\cm^3\s^{-1}$~\cite{HESS:HD2012}.
So far, none of those limits gets close to the expectation for vanilla WIMP models;
realistic models featuring particularly strong line signals, however, start to get constrained.

The small-scale ($\ell\gtrsim150$) gamma-ray \emph{anisotropies} observed by Fermi-LAT,
indicating the presence of some unresolved source population, can only
partially (if at all)  be explained by DM annihilation
\cite{Ackermann:2012uf}; on the other hand, this can already be used to
constrain the distribution of DM subhalos \cite{Fornasa:2012gu}.  While some
unidentified gamma-ray sources in principle qualify as candidates for
annihilating DM  \emph{clumps} 
\cite{Belikov:2011pu,Mirabal:2012em},
 the
presence of any unconventional sources in current gamma-ray data  seems very
unlikely after taking into account surveys at other wavelengths
\cite{Zechlin:2011kk,Zechlin:2012by}. 
 Finally, let us mention that the assumed
non-observation of  gamma-ray point sources from DM annihilation limits the
allowed abundance of \emph{ultracompact mini halos}
\cite{Ricotti:2009bs,Scott:2009tu,Lacki:2010zf}, 
which can be used to put extremely stringent constraints on the power of primordial
density perturbations~\cite{Bringmann:2011ut}.

\subsection{Signals}

Historically, there have been a couple of claims of potential DM signals in
gamma rays. For the GC, e.g., they correspond to DM masses in the
$\sim$MeV 
\cite{Boehm:2003bt,Hooper:2004qf,Picciotto:2004rp},
$\sim$10\,GeV 
\cite{Hooper:2010mq, Hooper:2011ti, Abazajian:2012pn}, $\sim$100\,GeV \cite{Cesarini:2003nr}, $\sim$TeV
\cite{bergstrom:2004cy,Horns:2004bk,Lin:2010fba,Dobler:2011mk} (sometimes with possible counterparts at radio
frequencies
  \cite{Finkbeiner:2003im,Finkbeiner:2004us,Dobler:2007wv}) 
  or even multi-TeV range 
 \cite{Profumo:2005xd,Belikov:2012ty,Cembranos:2012nj}. 
 Possible DM signals in gamma rays have also been claimed in the diffuse gamma-ray flux 
 \cite{deBoer:2005tm,Hutsi:2010ai}
  or from galaxy clusters \cite{Han:2012au}.
While it can be argued that there still remains a bit of controversy in some
of these cases, evidence is certainly not compelling; instead, in the past,
more refined analyses and new data often tended to either disfavor the DM hypotheses
previously put forward or make it less compelling in view of viable alternative, astrophysical 
explanations
\cite{Aharonian:2006wh,Bergstrom:2006tk,Stecker:2007xp,Weidenspointner:2008zz,Boyarsky:2010dr,Abdo:2010nz,Abazajian:2010zy,Han:2012uw,MaciasRamirez:2012mk}
(for a discussion of recent DM signal claims not only in gamma rays, see also Ref.~\cite{Bergstrom:2012fi}).
One reason for this is that the claimed signals typically rely on the presence
of some broad excess in the differential gamma-ray flux that was mostly
attributed to secondary photons which, as stressed before, makes the
identification of a DM signal intrinsically error-prone.\footnote{ 
The only exception is the 511\,keV line from $e^+e^-$ annihilation seen by Integral
\cite{Knodlseder:2005yq}. Its observed non-spherical distribution 
\cite{Weidenspointner:2008zz},
however, makes a DM interpretation highly unlikely -- which in any case would
be restricted to a very narrow mass range, $m_e\lesssim m_\chi\lesssim 3\,{\rm
MeV}$, in order not to overproduce continuum photons from final state
radiation \cite{Beacom:2004pe, Beacom:2005qv}.
}
In this context, it is also worth recalling that neither very small
($m_\chi\ll100\,$GeV) nor very large ($m_\chi\gg1\,$TeV) DM masses are easily
accommodated in realistic WIMP frameworks that successfully address
shortcomings of the standard model of particle physics.  Furthermore, one should appreciate the
fact that both CMB 
\cite{Cirelli:2009bb,Hutsi:2011vx,Galli:2011rz}
 and cosmic ray antiproton data \cite{Evoli:2011id}
provide very stringent constraints on the possibility of
$\mathcal{O}(10)\,$GeV WIMP DM -- even though such a possibility might be
 interesting from the point of view of direct DM searches
\cite{Bernabei:2008yi,Aalseth:2010vx,Aalseth:2011wp,Angloher:2011uu}.


The recently discovered hint for a monochromatic gamma-ray signal at around
130 GeV in the Fermi data of the GC region
\cite{Bringmann:2012vr, Weniger:2012tx}, on the other hand, would correspond
to a rather natural DM mass  of  $\mathcal{O}(100)$\,GeV and, even more
importantly, for the first  time provide evidence for a gamma-ray feature
which is widely regarded as a \emph{smoking gun signature} for DM
 \cite{Bergstrom:1997fj}.  Performing a spectral shape analysis in
target regions close to the GC (selected in a data-driven
approach by using photons at much lower energies), the signal was found to
correspond to a DM mass of $m_\chi=149\pm4\ ^{+8}_{-15}\GeV$ for an assumed
VIB signal \cite{Bringmann:2012vr} and $m_\chi=129.8\pm2.4\ ^{+7}_{-13}\GeV$
for a $\gamma\gamma$ line  \cite{Weniger:2012tx}, in each case with a local
(global) significance of almost 5 $\sigma$ (more than 3 $\sigma$). The deduced
annihilation rate depends on the DM profile; for an Einasto profile,
e.g., it is $\langle \sigma v\rangle_{\ell^+\ell^-\gamma}=(5.2\pm 1.3\
^{+0.8}_{-1.2})\times10^{-27}\cm^3\s^{-1}$ and $\langle \sigma
v\rangle_{\gamma\gamma}=(1.27\pm 0.32\ ^{+0.18}_{-0.28})
\times10^{-27}\cm^3\s^{-1}$, respectively (see Section \ref{sec:models} and 
Tab.~\ref{tab:line_limits} for a discussion of $\gamma Z$ and $\gamma H$ final states).  
This excess was  confirmed
independently \cite{Tempel:2012ey} by adopting a different statistical
technique based on kernel smoothing; this analysis also demonstrated that the
intrinsic signal width cannot be much larger than the energy resolution of
Fermi LAT ($\sim$10\% at 130 GeV) -- leaving only a VIB signal, a gamma-ray
line or a narrow box as possible explanation in terms of DM annihilation.
Later, Su \& Finkbeiner \cite{Su:2012ft} adopted a refined spatial template
analysis to demonstrate that the existence of a gamma-ray line emitting region
of radius $\sim$3$^\circ$ (see also Ref.~\cite{Tempel:2012ey}) 
close to the GC is preferred over
the no line hypothesis with a \emph{global} significance of more than
5$\sigma$ (under the assumption of an Einasto profile).

It is worth emphasizing that the above described signal is the \emph{only}
significant line-like feature in the sky that we currently find
distinguishable in the data from around 20\,GeV to at least 300\,GeV (we
checked this explicitly by performing line searches along the galactic disc, as well as a
subsampling analysis of anti-GC data -- see also Refs.~\cite{Bringmann:2012vr,
Weniger:2012tx, Su:2012ft}; conflicting claims~\cite{Tempel:2012ey,
Boyarsky:2012ca} may likely be explained as statistical fluctuations at the
expected level).  However, it is quite interesting to note that there might be
weak evidence for line signals, with much smaller significance but at the same
energy, in the direction of galaxy clusters \cite{Hektor:2012kc}. Further weak
evidence for such lines has also been found in some of the unassociated
gamma-ray point sources observed by Fermi \cite{Su:2012zg} (but see 
Ref.~\cite{Zechlin:2011kk,Zechlin:2012by}).
 These 
indications are currently intensely debated 
\cite{Hooper:2012qc, Mirabal:2012za, Hektor:2012jc}
 -- 
if eventually confirmed with better statistics than currently available, they would 
significantly strengthen a DM interpretation. Let us also
mention that there is no correlation between the Fermi
bubbles~\cite{Su:2010qj, Dobler:2011mk} and the line signal at the GC
 \cite{Tempel:2012ey,Bergstrom:2012fi, Su:2012ft}; the significant overlap
\cite{Profumo:2012tr} of the bubbles with the target regions adopted in
Refs.~\cite{Bringmann:2012vr, Weniger:2012tx} is thus purely accidental and
related to the peculiar angular distribution of signal and background photons
(see also Ref.~\cite{Nezri:2012xu}).

\begin{figure}[t]
  \begin{center}
    \includegraphics[width=0.7\textwidth]{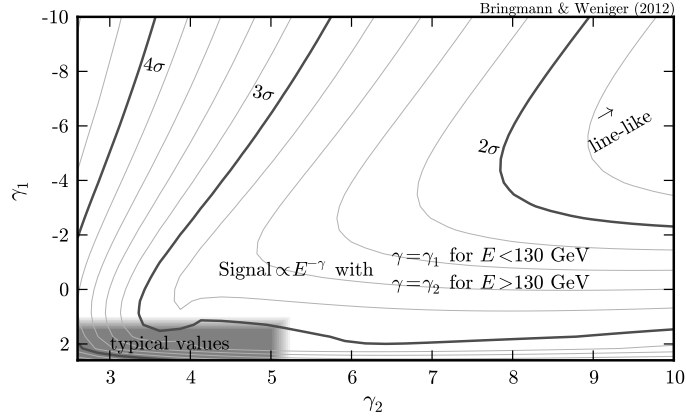}
  \end{center}
  \caption{This plot shows the confidence contours obtained when fitting the
  130 GeV signature with a broken power-law with spectral break from
  $\gamma_1$ to $\gamma_2$ at 130 GeV (plus the usual background power-law
  with free normalization and slope). Best fits are obtained for
  $\Delta\gamma=\gamma_2  -\gamma_1 \gg 10$, when the signal approaches a
  line-like shape; the gray area indicates parameters that are realistically
  accessible for astrophysical sources (see e.g.~\cite{:2010yg}). In light of
  these results, it is not surprising that also the fit by a hard power-law
  with superexponential cutoff -- as e.g.~realized for pulsar emission --
  plus a power-law background is disfavored w.r.t.~a monochromatic line by at
  least $3\sigma$.}
  \label{fig:powerlaw}
\end{figure}

As already stressed, the \emph{intrinsic signal width} is small: assuming a Gaussian
instead of a monochromatic signal, we find an upper limit of 18\% at 95\%CL (though a \emph{pair} 
of lines might provide a marginally better fit, see Section \ref{sec:models}). 
A toy example for an extremely sharp gamma-ray feature with astrophysical
 origin would e.g.~be ICS emission from a hypothetical nearly
 monochromatic $e^\pm$ population at the GC (see e.g.~Ref.~\cite{Weniger:2012ms}). Such a population might arise
 from pile up of electrons during synchrotron cooling, but the resulting ICS
 gamma-ray spectrum is still disfavoured w.r.t.~a monochromatic line by about
 $3\sigma$.
In
Fig.~\ref{fig:powerlaw}, we demonstrate furthermore explicitly that a broken
power-law, as suggested in Ref.~\cite{Profumo:2012tr}, does \emph{not} provide
a reasonable fit to the data unless one allows for an absurdly large spectral
break $\Delta\gamma\equiv\gamma_2-\gamma_1\gg 10$; the best fit is obtained
in the line-like limit $\Delta\gamma\to\infty$. As also indicated in the
figure, the required spectral indices both above ($\gamma_2$) and below ($\gamma_1$) 
the break would in
that case be well outside the range of values observed in standard
astrophysical sources (the power-law background in our fit, on the other
hand, has a spectral index consistent with the expected value of $\simeq2.6$~\cite{FermiLAT:2012aa}, mostly determined by  cosmic-ray proton collisions with the interstellar
medium).\footnote{To generate the plot, we redid the analysis from
Ref.~\cite{Weniger:2012tx} in Reg4 (SOURCE events), replacing the
monochromatic line by a broken power-law that changes its spectral
index from $\gamma_1$ to $\gamma_2$ at 130 GeV.}
Note that a smooth change of 
$\gamma$ would make the fit quality even worse, so similar conclusions 
hold for the more commonly encountered 
case of a power-law with a super-exponential cutoff, i.e.~$dN/dE\propto
E^{-\gamma} \exp[-(E/E_{\rm cut})^a]$: for 'typical' values of $\gamma$ and
$a$ (roughly
$1\lesssim\gamma,a\lesssim2$
\cite{Abdo:2009ax,1995ApSS.230..299N}),
 but free
$E_{\rm cut}$, we find that such a spectrum is always disfavored w.r.t.~a
monochromatic line by at least $3\sigma$.

Let us now briefly turn to what currently appears as the greatest challenge to
a DM interpretation of the observed signal (of course, this necessarily
reflects our own bias; for extended discussions see e.g.~Refs.~\cite{
Bringmann:2012vr, Weniger:2012tx, Su:2012ft, Hektor:2012ev}; a quite thorough
discussion of possible instrumental effects can be found in 
Ref.~\cite{Finkbeiner:2012ez}). The first caveat is
that the signal center seems to be displaced, by about $1.5^\circ$ or 200\,pc,
from the dynamical center of the galaxy \cite{Su:2012ft} (see
Ref.~\cite{Tempel:2012ey} for an early indication); even if such a displacement 
might in principle be possible, following our discussion in Section \ref{sec:GC}, this 
certainly came as a surprise.
While the \emph{most likely} center of the 
emission is clearly displaced by an amount as stated above, on the other hand, the photon distribution still seems to be statistically \emph{consistent}
with a single source -- with an NFW or Einasto profile -- centered exactly at the GC \cite{Rao:2012fh} (see also Ref.~\cite{Yang:2012ha} for a corresponding earlier claim).
An even stronger threat to the DM hypothesis might thus be the indication, so far
at a weaker level of significance, for a line in part of the gamma rays from
cosmic-ray induced air showers in Earth's atmosphere (commonly referred to as
Earth limb or Earth albedo)~\cite{Su:2012ft, Finkbeiner:2012ez}. However, the 
problematic limb photons only appear at a very specific range,
$30^\circ\lesssim\theta\lesssim45^\circ$,  of incidence angles
(unlike the signal from the
GC which shows up at all $\theta$). Furthermore, the
majority of events with these incidence angles come actually \emph{not} from
the limb but are of astrophysical origin and this larger sample does not show any
evidence for a 130 GeV feature~\cite{Finkbeiner:2012ez, Hektor:2012ev}.  This
confusing situation
might well be an indication that the limb excess is merely a statistical fluctuation
that soon will disappear with more limb data.

So far no compelling alternative instrumental \cite{Whiteson:2012hr,
Finkbeiner:2012ez} or astrophysical mechanism has
been proposed that could actually produce such a line at 130 GeV (see 
Ref.~\cite{Aharonian:2012cs} for an interesting proof-of-principle with
fine-tuned pulsar winds -- which however cannot explain the extended
morphology of the signal). We stress that all analyses of the line
signal so far rely on the publicly available Fermi data and information only,
and that line searches operate by construction at the statistical and
systematical limitations of the instrument.
There has not yet been an official statement from the Fermi collaboration concerning 
the signal, in particular with respect to whether
the energy reconstruction of Pass 7 events is reliable at energies above $100\GeV$ 
in light of the recent findings.\footnote{
Note that the latest
official compilation of Fermi line limits \cite{Ackermann:2012qk} was
finalized \emph{before} the first \cite{Bringmann:2012vr} indication for the
line signal was announced; it relies on 24 rather than 43 months of
data and takes a significantly larger ROI.
The tentative signal claim is thus not in tension with those limits \cite{Weniger:2012tx}.
} 
Eventually,
such an independent confirmation of the 130 GeV excess will of course be indispensable.

\section{What could we learn from a signal?}
\label{sec:implications}

Gamma rays may carry important and nontrivial information
about the nature of the DM particles. Let us now demonstrate in more detail
what kind of information could  actually be extracted in case of
a signal identification, in particular in case of a sharp spectral signature. 
For definiteness, we will take the tentative line
signal as an \emph{example} and assume in this Section that it can indeed be explained by DM.

\begin{figure}[t]
  \begin{center}
    \includegraphics[width=0.7\textwidth]{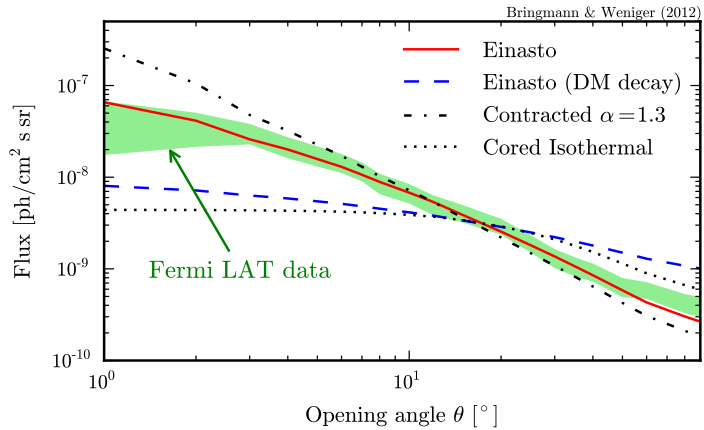}
  \end{center}
  \caption{\label{fig:profile} Comparison of different flux profiles as function
  of the opening angle $\theta$ of an hourglass-shaped ROI that is centered on the GC (see
  text for detailed definition). In green we show the $\pm1\sigma$ uncertainty
  band of the line flux measured inside this region by Fermi LAT after 3.6 years:
  while compatible with a standard Einasto profile at $\theta \gtrsim1^\circ$
  (as well as an NFW profile; see text), it is incompatible with both a
  cored and a sufficiently contracted profile, as well as with a signal from
  DM decay. 
  The green bars indicate which values of $\theta$ we actually use in the
  fits; the profiles are arbitrarily normalized such that they reproduce the
  correct flux for $\theta=20^\circ$. Note that we do not make any assumption
  about a possible displacement of the signal and that the ROI is centered on
  the GC.}
\end{figure}

\subsection{Dark Matter distribution}
\label{sec:signal_distribution}

A gamma-ray line would allow to study the distribution of DM in the GC with
unprecedented accuracy, which  could serve as important feedback for 
state-of-the-art numerical simulations of gravitational clustering. 
To illustrate this, we show in  Fig.~\ref{fig:profile} 
the $\pm1\sigma$ range of the line flux as function of the opening angle 
$\theta$ of a ROI
centered on the GC (green band).\footnote{
We use here an ROI with hourglass-like shape, defined by
$\psi\!<\!\min(3^\circ, \theta)$ plus $\psi\!<\!\theta$ and $|b/\ell|\!>\!0.7$
($\psi$ is the angular distance from the GC). Otherwise, we use the
same procedure as in Ref.~\cite{Weniger:2012tx} to obtain the line flux,
i.e.~a power-law + line fit to SOURCE class events.
}
  Obviously, for large
galactocentric distances, the flux drops because the GC signal is washed out. 
Remarkably, the flux profile is perfectly consistent with the 
predictions from a standard Einasto profile (red solid line). The same is true
for an NFW profile (not shown), which would in this context only significantly
differ closer to the GC, at angles $\theta\lesssim1^\circ$. 
The
dash-dotted black line shows for comparison the prediction for a DM
annihilation signal from a sufficiently contracted profile (chosen to be essentially 
equivalent, for the angular resolution of Fermi-LAT, to a point source at the GC); 
at angles $\theta\lesssim5^\circ$ the measured flux starts to deviate from the
predictions, indicating that the 130 GeV signal is not a point source but
extended up to these angles. On the other hand, the signal is too concentrated
to be compatible with a cored profile (dotted black line, here for a core 
radius of $3.5\kpc$). Note that a contribution from Milky
Way subhalos could further boost the signal at angles $\theta\gtrsim 20^\circ$
by a factor of a few~\cite{Pieri:2009je}, which however is not observed
and thus might be used to place constraints on subhalo models. DM decay
(blue dashed line) would also lead to a flux that is too weak at the GC to
be compatible with the observations (unless extreme assumptions on the profile
are made~\cite{Buchmuller:2012rc}, which however are likely to be in conflict with
microlensing and dynamical constraints~\cite{Iocco:2011jz}). 

 As already mentioned, there is some evidence for a $\sim1.5\,^{\circ}$ offset of the 
 signal with respect to the GC \cite{Su:2012ft} (though a centered signal
 might also be consistent with the data \cite{Rao:2012fh}). While this observation was certainly 
 unexpected, it might possibly be explained by the interplay between baryons forming 
 a bar and DM~\cite{kuhlen}. The same bar would however also likely destroy a DM
 cusp in the center~\cite{Weinberg:2001gm}; tidal disruption by the central supermassive
 black hole may be a further issue to worry about. More data, as well as more
detailed simulations along the lines of~\cite{Guedes:2011ux} are thus needed to
settle in how far the morphology of the 130 GeV feature is compatible with
theoretical expectations. In fact, such an improved understanding
could eventually allow  to infer important details about the formation history of our Galaxy.

Once the GC signal is established, an exciting future application would be a precise 
all-sky survey to look
 for the same 130\,GeV feature, aiming at a (partial) map of the 
Galactic and cosmological DM distribution. For sufficiently large substructure boosts, the
130\,GeV feature could for example appear as a bump in the IGRB or in the gamma-ray 
spectrum of galaxy clusters. So far, no corresponding lines were found in the IGRB  \cite{Abdo:2010dk}, 
which would already put the more optimistic models for substructure evolution from Ref.~\cite{Zavala:2009zr}  under some tension if  the small-scale cutoff in the subhalo distribution is roughly $10^{-6}M_\odot$ or less (while the reference model used for the Fermi-LAT analysis \cite{Abdo:2010dk} leads to constraints almost two orders of magnitude weaker 
than the GC signal strength).
On the other hand, there \emph{is} a possible weak 
indication from galaxy clusters~\cite{Hektor:2012kc}.  If confirmed, it would 
necessarily imply a rather small value for the small-scale cutoff in the subhalo distribution
in order to produce the required large boost factors of $\mathcal{O}(10^3)$; this, in turn, could 
be used to obtain highly complementary bounds on the underlying WIMP model. 
There is presently no sign for a 130\,GeV line in dwarf  galaxies and the resulting limits
on $(\sigma v)_{\gamma\gamma}$ are about one order of magnitude weaker than what
is needed to explain the GC signal \cite{GeringerSameth:2012sr}; however, these limits are significantly affected by uncertainties in the DM distribution in the dwarfs and in the most  favorable case a signal might appear already after a few times larger exposure than currently collected
by Fermi-LAT. 
With more data, it may also be possible to identify a few individual DM subhalos 
(see e.g.~the heavily disputed weak indications for a 130 GeV signal
from unidentified sources of the 2FGL \cite{Su:2012zg,Hooper:2012qc,Mirabal:2012za, Hektor:2012jc}).
The galactic distribution and total number of those subhalos would in principle provide invaluable information on the DM
distribution and allow to further discriminate between subhalo models that are currently discussed.
Note, however, that presently no precise estimate exists of how many subhalos are actually \emph{expected} to be visible, in the light of  results from $N$-body simulations, if the spectrum is dominated by a line; for a secondary spectrum (assuming $\bar b b$ annihilation and $m_\chi\simeq$100\,GeV), on the other hand, it was predicted that Fermi should have seen up to a few subhalos \cite{Pieri:2007ir,Kuhlen:2008aw} and the non-observation places a limit on the annihilation cross section
comparable to the one obtained from the IGRB \cite{Ackermann:2012nb}. More detailed
future studies in this direction would thus certainly be both very interesting and worthwile.


\subsection{Dark Matter models}
\label{sec:models}


At the time of this writing, the literature has already seen a considerable
amount of model-building efforts to explain the line signal in terms of
\emph{annihilating DM}. This ranges from
phenomenological and in some sense model-independent approaches
\cite{Buckley:2012ws,Tulin:2012uq}, or analyses in the context of effective
field theories  \cite{Rajaraman:2012db,Kang:2012bq,Frandsen:2012db}, to
concrete model building.  Proposed solutions that mostly fall into the latter
category include an additional  $U(1)$ symmetry \cite{Dudas:2012pb}, DM as the
lightest state of a new scalar multiplet 
\cite{Cline:2012nw,Fan:2012gr},
 right-handed
sneutrino \cite{Choi:2012ap} or neutrino \cite{Bergstrom:2012bd} DM, axion-mediated DM annihilation
\cite{Lee:2012bq,Lee:2012wz}, two-component DM  \cite{Acharya:2012dz}, magnetic inelastic
DM 
\cite{Weiner:2012cb,Weiner:2012gm},
 dipole-interacting DM 
\cite{Heo:2012dk,Cline:2012bz},
as well as 
scalar DM in extensions of the Higgs triplet \cite{Wang:2012ts} or Zee-Babu model \cite{Baek:2012ub}. Even
neutralinos have been proposed as a possible cause of the signal, albeit in
non-minimal versions of supersymmetry like no-scale $\mathcal{F}$--$SU(5)$
\cite{Li:2012tr,Li:2012jf,Li:2012mr}
 or the NMSSM \cite{Das:2012ys,Lee:2012bq,Kang:2012bq} -- each
time, however, arguing for additional indications in favor of the respective
model in collider data. The possibility of \emph{decaying DM} being
responsible for the signal has also been entertained
\cite{Kyae:2012vi,Buchmuller:2012rc,Park:2012xq} -- though the expected angular
dependence of the signal in this case is hardly consistent with observations, see the previous subsection.

One generic problem for any realistic model-building is that the annihilation
cross section required to fit the data is considerably larger than typically
expected for thermally produced DM, at least if the relic density is set by
the tree-level annihilation rate. On the particle physics side, possible ways
to enhance the annihilation rate  in that case include the Sommerfeld
enhancement 
\cite{sommerfeld,Hisano:2003ec,Hisano:2004ds,ArkaniHamed:2008qn}
in the presence of new light bosonic messenger particles that mediate an
attractive force between the initial state DM particles or, at the cost of
some fine- tuning, the presence of a resonance (i.e.~$s$-channel annihilation
via a new neutral particle with $m\simeq2m_\chi$, the same spin and $CP$ 
properties as the initial state);
yet another mechanism might be cascade annihilation \cite{Bai:2012qy}. 
 On the astrophysical side,  larger annihilation 
fluxes arise by adopting a larger
local DM density for the profile normalization or a profile that is steeper in 
the innermost part than
our reference Einasto profile, Eq.~(\ref{einasto}); see, however,
Fig.~\ref{fig:profile} for the relatively tight constraints on the latter
option.  Note that even if one relaxes the theoretically appealing
assumption of thermal DM production, one needs to worry about large
annihilation rates at tree-level because they would produce secondary photons
potentially in stark conflict with continuum gamma-ray data
\cite{Buckley:2012ws,Buchmuller:2012rc,Cohen:2012me,Cholis:2012fb,
Huang:2012yf}; also antiproton \cite{Buchmuller:2012rc,Asano:2012zv} and radio
\cite{Asano:2012zv,Laha:2012fg} data are quite efficient in constraining such large
annihilation rates. This fact can be used to rule out e.g.~Wino or Higgsino DM
as an explanation for the line (see also below). Antiprotons could
be constraining for future experiments not only because of the tree-level
annihilation rate, but also due to the associated DM annihilation into $gg$
final states if the $\gamma\gamma$ signal is dominated by colored particles in
the loop \cite{Chu:2012qy}.

\begin{figure}[t]
  \begin{center}
    \includegraphics[width=0.7\textwidth]{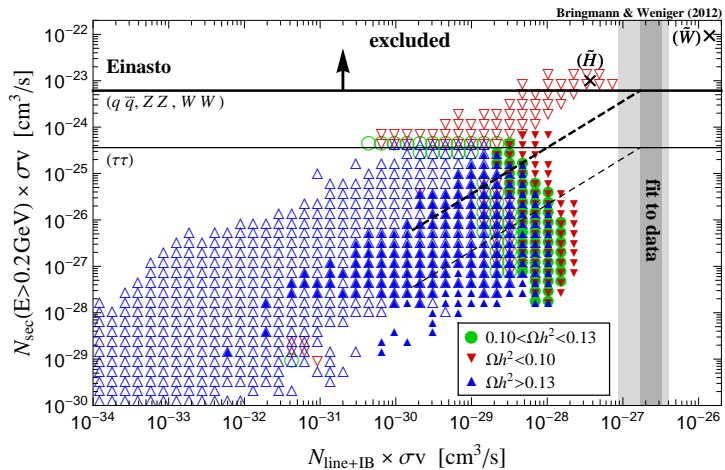}
  \end{center}
  \caption{\label{fig:SUSY_scan} SUSY scan  comparing the expected number of
  quasi-monochromatic photons  ($120\,{\rm GeV}\leq
  E_\gamma\leq140\,{\rm GeV}$) to the number of secondary photons. Green
  (red, blue) points correspond to models where thermal production leads to a
  relic density in (smaller than, larger than) the observed range.  Filled
  symbols indicate models where VIB contributes at least 3 times more photons
  than $\gamma\gamma$ and $\gamma Z$. Exclusion limits \cite{Cholis:2012fb}   and signal fit (at $1\sigma$ and $2\sigma$) both assume an
  Einasto profile. Dashed lines show the effect of enhancing the annihilation flux
  by the same amount in both exclusion ($|b|,|l|<5^\circ$) and signal ROI; only models below the dashed line may thus in
  principle  explain the line at 130\,GeV.}
\end{figure}

In order to illustrate the above point, we consider in
Fig.~\ref{fig:SUSY_scan} the result of a large scan (for details, see
Ref.~\cite{Bergstrom:2010gh}) over the  parameter space of the cMSSM and a
phenomenological MSSM-7, keeping only neutralino DM models where IB,
$\gamma\gamma$ and $\gamma Z$ photons for $E_\gamma\in[120,140]$\,GeV dominate
the secondary contribution by a factor of at least  5; for reference, we also show the 
case of pure Wino (`$\tilde W$') and Higgsino (`$\tilde H$') DM. Filled symbols
correspond to models where IB photons outnumber line photons by at least a
factor of 3. Assuming an Einasto profile as in Eq.~(\ref{einasto}), we also
show the required signal strength to account for the line observation (shaded
area) as well as limits  \cite{Cholis:2012fb} on the continuum flux from the
GC region (solid lines). Note that both limits and signal region are roughly
proportional to $\left(\int d\Omega ds\, \rho_\chi^2\right)^{-1}$, albeit
integrated over slightly different regions near the GC; assuming this factor
to be the same for both regions, dashed lines indicate  how limits and signal
region would change if adopting a profile that is different from
Eq.~(\ref{einasto}).  From this figure, we can draw at least three important
conclusions: i)~As already anticipated, the large annihilation rate required
to explain the signal cannot easily be achieved for thermally produced
neutralino DM. ii) Even when enhancing the annihilation rate such as to
sufficiently increase the production of $\gamma\gamma$ or $\gamma Z$ final
states (e.g.~by a higher central DM density), 
it is very difficult to do so without violating the bounds from
continuum gamma rays.
In fact, the observed correlation
between loop- and tree-level rates is generally expected from the optical
theorem and should thus not only apply to neutralino DM \cite{Asano:2012zv}.
iii) VIB, on the other hand, does not follow this pattern and can thus be
argued to be a more natural explanation for such a strong line-like signal
 -- still in need, however, of $\mathcal{O}(10)$ enhancement factors for
standard\footnote{ 
Let us stress that all supersymmetric models of Fig.~\ref{fig:SUSY_scan} 
assume unification of gauge couplings at the GUT scale, which
prohibits a large Wino fraction of the lightest neutralino.  Scanning a simple
phenomenological MSSM-9 \cite{Bergstrom:2008gr}, where this assumption
is relaxed,  we found the top right branch of non-thermally produced models 
(open red symbols in Fig.~\ref{fig:SUSY_scan}) to extend all the way to the 
pure Wino case -- albeit always above the (extended)  dashed line. However, 
we could not find any models with larger IB rates than shown in 
Fig.~\ref{fig:SUSY_scan} (possibly due to the restricted nature of the scan and 
MSSM version employed).\\ 
Secondary photons from electroweak and strong corrections, see Section 
\ref{sec:spec}, were not included and would move the VIB dominated 
models  in Fig.~\ref{fig:SUSY_scan} somewhat upwards. While a dedicated 
future analysis is certainly  warranted, let us stress that we still expect the 
continuum gamma-ray limits to be easily satisfied for VIB dominated, thermally produced 
neutralinos (anti\-proton constraints from these channels \cite{Garny:2011ii} 
are likely less stringent  \cite{Bringmann:2012vr}).} 
neutralino DM.
We note that such an enhancement may actually not be unrealistic given the 
significantly larger values of the local DM density $\rho_\odot$ that are found
when assuming a non-spherical DM profile or the presence of a dark disc
\cite{Garbari:2012ff}; furthermore, the most recent simulations of Milky Way like 
galaxies suggest that baryons should increase the DM density in the central 
parts by a factor of almost 3, in a way compatible with the angular distribution 
of the signal as shown in Fig.~\ref{fig:profile}  \cite{kuhlen}.
While most DM model-building so far has focussed on an explanation of the 130\,GeV 
feature in terms of monochromatic gamma-ray lines, also VIB-dominated signals
have been considered explicitly in this context \cite{Bringmann:2012vr,Bergstrom:2012bd,Shakya:2012fj}.

\begin{figure}[t!]
  \begin{center}
    \includegraphics[width=0.7\textwidth]{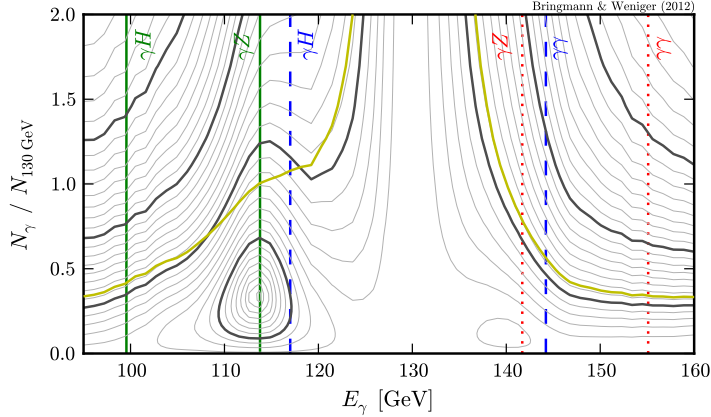}
  \end{center}
  \caption{Significance contour (thick black lines in $1\sigma$ steps) and
  upper limits (yellow line; $95\%$CL) for a second line. Assuming that the
  130 GeV feature is due to $\gamma\gamma$ (green solid), $\gamma Z$ (blue
  dashed) or $\gamma H$ (red dotted), the vertical lines show the
  corresponding positions of the other two lines. A very weak hint for a
  $\gamma Z$ line at $114\GeV$ can be identified.}
  \label{fig:gggZ}
\end{figure}

The possibly only way to avoid the above considerations may be to strongly
restrict any coupling of DM to charged standard model particles lighter than
$m_\chi$ \cite{Asano:2012zv}: loop signals could then easily dominate over VIB
signals without being in conflict with constraints arising from tree-level
annihilations. In such a case, one would rather generically expect not only
one but at least two lines \cite{Rajaraman:2012db} and the observed ratio of
photon counts (or limits on those) can provide crucial information about the
underlying particle model \cite{Rajaraman:2012db, Oda:2012fy}. In
Fig.~\ref{fig:gggZ} we therefore provide significance contours and upper
limits for a second line besides the observed 130\,GeV feature; for
convenience, we summarize these results in Tab.~\ref{tab:line_limits} in terms
of limits on the annihilation cross section $(\sigma v)_{\gamma X}$ under the
assumption that the signal corresponds to DM annihilation into $\gamma Y$ 
(for $X, Y = \gamma, Z, h$). Interestingly, as observed earlier
\cite{Rajaraman:2012db, Su:2012ft}, one can see a weak indication 
(with a significance of around 1.4\,$\sigma$) for a second
line at 114\,GeV -- which coincides surprisingly well with the energy expected
for a $\gamma Z$ line if the 130\,GeV feature can be attributed to DM
annihilation into $\gamma\gamma$; for this case, we also state the best fit
value for the ratio of cross sections.

\begin{table}[t!]
  \centering
  \begin{tabular}{ccccccc}
    \toprule
    $\gamma X$  && $m_\chi$ [GeV] &
    $\langle\sigma v\rangle_{\gamma X}$ [$10^{-27}$cm$^3$s$^{-1}$] &
    $\frac{\langle\sigma v\rangle_{\gamma\gamma}}{\langle\sigma v\rangle_{\gamma X}}$&
    $\frac{\langle\sigma v\rangle_{\gamma Z}}{\langle\sigma v\rangle_{\gamma X}}$&
    $\frac{\langle\sigma v\rangle_{\gamma H}}{\langle\sigma v\rangle_{\gamma X}}$ \\
    \midrule
    $\gamma\gamma$ && $129.8\pm2.4^{+7}_{-14}$ & $1.27 \pm 0.32 ^{+0.18}_{-0.28}$ &
    1  & $0.66^{+0.71}_{-0.48}$ & $<0.83$ \\[3pt]
    $\gamma Z$     && $144.2\pm2.2^{+6}_{-12}$ & $3.14 \pm 0.79 ^{+0.40}_{-0.60}$ &
    $<0.28$ & 1 &  $<1.08$   \\[3pt]
    $\gamma H$     && $155.1\pm2.1^{+6}_{-11}$ & $3.63 \pm 0.91 ^{+0.45}_{-0.63}$&
    $<0.17$ & $<0.79$ & 1  \\
    \bottomrule
  \end{tabular}
  \caption{\label{tab:line_limits} Upper limits at 95\%CL (or best-fit value
  with $\pm1\sigma$ error) on the branching ratios into the secondary line,
  assuming that the primary line at $E_\gamma=130\GeV$ is due to annihilation
  into $\gamma X$ with $X=\gamma$, $Z$ or $H$. Note that $\langle \sigma v
  \rangle_{\gamma Z}/\langle\sigma v\rangle_{\gamma\gamma}<2.01$ at 95\%CL.}
\end{table}

\section{Future prospects}
\label{sec:prospects}

\subsection{Next decade}
\label{sec:near_future}
The next ten years will bring a plethora of new results in indirect DM
searches.  It is right now that experiments start to probe vanilla WIMP DM
models and thus will either identify a signal or exclude many of the most common scenarios.
Ongoing experiments like Fermi-LAT, HESS-II, VERITAS and MAGIC will continue
to take data, may identify new targets for DM searches, profit from a better
understanding of astrophysical backgrounds and prepare the stage for planned
 instruments like CTA or GAMMA-400 with considerably improved characteristics
for DM searches. Indirect detection with gamma rays will also profit from an
interplay with upcoming results from neutrino searches with IceCube,
anti-matter searches with AMS-02, results from the LHC as well as from next-generation
direct WIMP detectors.  Furthermore, continuously improving
results from N-body simulations that realistically take  into account the
various components of baryonic matter will sharpen our understanding of the
signal morphology.

Assuming a ten year lifetime of Fermi-LAT, the limits on the annihilation
cross-section that were derived from observations of nearby dwarf galaxies
with 2 years of data~\cite{Ackermann:2011wa} would improve on purely
statistical grounds by a factor of $\sqrt{5}$ to $5$, depending on the
annihilation channel and the DM mass (which determines whether the limits are derived in
the signal-dominated high-energy regime or in the low-energy regime dominated by
the diffuse gamma-ray background). Optical surveys like Pan-STARRS~\cite{Kaiser:2002zz}, the
Dark Energy Survey~\cite{Abbott:2005bi} or the Stromlo Missing Satellite
Survey~\cite{stromloWebsite}
could increase the 
number of known dwarf spheroidals by a
factor of 3, which could additionally increase the constraining power by a
factor of $\sqrt{3}$ to $3$ in the most optimistic case
\cite{Tollerud:2008ze, Cotta:2011pm,stromlo}.  
Further significant
improvements are expected from the upcoming Pass 8 version of the LAT event
reconstruction, which will lead to an enhanced effective area for
high energy gamma rays, better hadron rejection
 and an improved energy resolution~\cite{pass8}. It is
hence conceivable that Fermi-LAT dwarf limits will improve by a factor up to
ten, which could allow to constrain WIMPs with thermal annihilation rate
into $\bar bb$ up to DM masses of $\sim600\GeV$.  Similar improvements might
be expected for limits from galaxy clusters \cite{Ackermann:2010rg,
Zimmer:2011vy}. DM searches in the GC \cite{Cohen:2012me,
Cholis:2012fb, Hooper:2012sr} and the halo \cite{Zaharijas:2010ca}, on the other
hand, will 
mostly profit from a refined understanding of astrophysical backgrounds;
results from different groups are expected soon (already now
e.g.~Ref.~\cite{Hooper:2012sr} finds limits that severely constrain thermal
cross-sections into $\bar{b}b$ for $\lesssim100\GeV$ DM masses, though with
significant dependence on the details of the halo and background model).
%
The best limits on annihilation into gamma-ray lines \cite{Weniger:2012tx} or
VIB features  \cite{Bringmann:2012vr} are right now based on
almost four years of data; more data and improved event
reconstruction will strengthen them by at least a factor of $\sqrt{3}$.
 
By now, HESS observations of the GC provide the strongest limits on DM
annihilation with $m_\chi\gtrsim 700\GeV$ \cite{Abramowski:2011hc}, down to
annihilation cross-sections of $4\times 10^{-25}\cm^3\s^{-1}$ in case of
$\bar{b}b$ final states. The
newly mounted 28m-diameter telescope HESS-II has an about 3 times lower energy
threshold than HESS-I, as well as additional timing information which will
improve the rejection of cosmic rays~\cite{Hess2:Hd2012}. Both improvements
will help to extend the current constraints to lower masses and values of the
annihilation cross-section.
If Fermi-LAT identifies a DM signal candidate at
high enough energies, HESS could quickly confirm it thanks to its large
effective area, and provide additional information about the variability and
spatial extent of the source. Further results are anticipated from VERITAS,
MAGIC and AMS-02 as well.
Since VERITAS and MAGIC see the GC at most at angles $\sim33^\circ$ above
horizon (while HESS at angles up to $84^\circ$), however, they are mainly interesting
for observations of dwarf galaxies and less for DM searches at the GC.
Substantial improvement in the TeV regime should eventually come with CTA.
Following Ref.~\cite{CTADMprospects}, observations of the GC are expected to
exclude cross-sections down to the thermal one at TeV DM masses, which would
be an improvement of up to an order of magnitude with respect to the current
HESS constraints (note, however, that  Ref.~\cite{CTADMprospects} adopts a factor 
of $\sim$\,10
substructure boost of the GC signal w.r.t.~what is
expected from standard smooth Einasto or NFW profiles).
Further improvements with 
respect to HESS or VERITAS are also expected for
dwarf galaxy observations, although they would still hardly be competitive
with results from space-based instruments.  



Future space-based instruments like GAMMA-400 or CALET/DAMPE will -- thanks to
an extended imaging calorimeter and a large lever arm to the converter
foils -- have a much better energy and angular resolution than Fermi LAT.
However, they will come with a somewhat smaller effective area. For that
reason, DM limits from e.g.~dwarf spheroidal observations would likely only be
somewhat strengthened in the background limited regime at low energies.
On the other hand, these instruments would be excellent machines for
detailed follow up studies of DM signal candidates that might be identified in
the Fermi-LAT data, in particular in case of pronounced spectral features.

\begin{figure}[t]
  \begin{center}
  \includegraphics[width=0.95\textwidth]{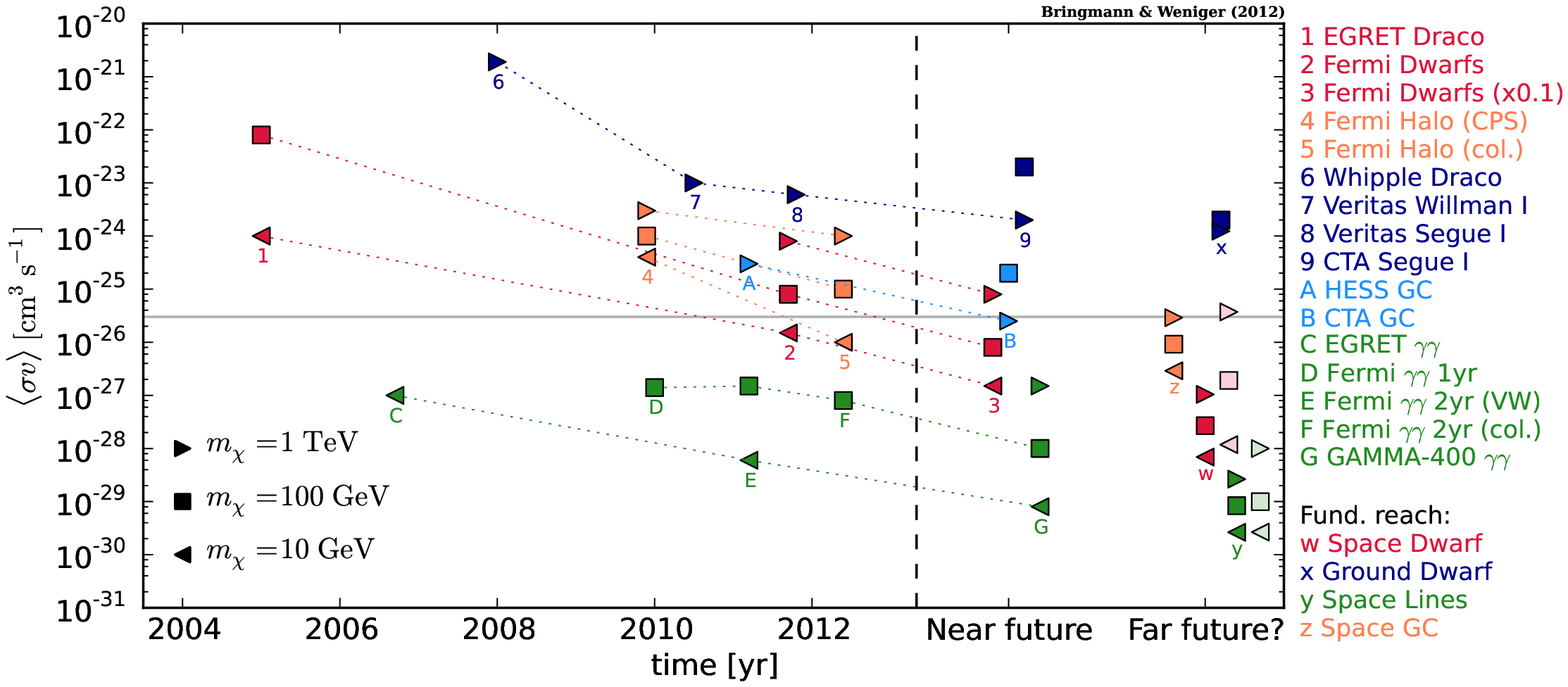}
  \end{center}
  \caption{\label{fig:time}Time evolution of limits. References: EGRET
  Draco~\cite{Bergstrom:2005qk}; Fermi Dwarfs~\cite{Ackermann:2011wa}; Fermi
  Halo (CPS)~\cite{Cirelli:2009dv}; Fermi Halo (col.)~\cite{Ackermann:2012rg};
  Whipple Draco~\cite{Wood:2008hx}; Veritas Willman I~\cite{Acciari:2010ab};
  Veritas Segue I~\cite{Vivier:2011sm}; 
  CTA Segue~I and GC~\cite{CTADMprospects}; 
  HESS GC~\cite{Abramowski:2011hc}; EGRET
  $\gamma\gamma$~\cite{Pullen:2006sy}; Fermi $\gamma\gamma$
  1yr~\cite{Abdo:2010nc}; Fermi $\gamma\gamma$ 2yr
  (VW)~\cite{Vertongen:2011mu}; Fermi $\gamma\gamma$ 2yr
  (col.)~\cite{Ackermann:2012qk}; GAMMA-400~\cite{Bergstrom:2012vd}.  For the
  dark red, dark green, blue and orange far future `fundamental' limits, we took only
  into account
  systematic limitations (basically assuming that all relevant systematics can
  be understood at the $1\%$ level); the corresponding observational times can
  be extremely large in case of space-based telescopes, but are realistic for
  IACTs. For comparison, the light green and light red symbols show the limits
  obtained for hypothetical sky exposures about 100 times larger than 10 years
  Fermi LAT observations in survey mode.}
\end{figure}

In Fig.~\ref{fig:time} we provide a convenient
\emph{summary plot} of limits on the DM annihilation cross section into $\bar bb$ (in red
and blue for space- and ground-based instruments, respectively) as well as in
$\gamma\gamma$ (in green).
The limits are collected
as a representative selection of different instruments, and we concentrate on
observations of dwarf spheroidal galaxies, the GC and the
Galactic halo, since they provide right now the most stringent constraints.
All limits are shown as function of the time of their publication, and were
derived for common assumptions on the DM profile (NFW profiles in
most cases; in case of e.g.~GC and Galactic halo limits the adopted $J$-values
are mutually consistent within a factor of $\sim2$ [except for CTA, see above]);
the different symbols correspond to limits  for DM masses
of $10$, $100$ and $1000\GeV$. To the right of the dashed black line, limits
expected for the next decade are shown, as well as limits that might be
achievable in the more distant future (to be 
discussed in Sec.~\ref{sec:fundamental}). In the past eight years, most limits
have improved by an order of magnitude; a similar improvement is expected
during the upcoming ten years. Even without excessive boost factors, these
limits start to reach deep into the parameter space of WIMP DM
models. In particular observations of the GC with Cherenkov
telescopes at high energies as well as observations of dwarf spheroidal
galaxies with space-based instruments at lower energies have a great potential
for deriving constraints or discovering a signal.

Finally, the prospects for a further study of
\emph{the 130 GeV feature} in the Fermi-LAT data, if it persists, are extremely good.
HESS-II has just seen its first light and given the good performance foreseen for
the instrument in hybrid mode, it should allow a quick confirmation of the
signal reported in \cite{Bringmann:2012vr,Weniger:2012tx}, if systematic uncertainties are
sufficiently under control~\cite{Bergstrom:2012vd}; for CTA, a mere 50 hrs
of data might be enough to confirm the signal \cite{Aleksic:2012cp}. In the case of the future
space-based telescopes GAMMA-400 and DAMPE, the improved energy resolution
provides an enormous potential not only for the detection of the 130 GeV
feature, but also for the efficient discrimination of a VIB feature from one or several
lines \cite{Bergstrom:2012vd, Li:2012qg}.
However, even with GAMMA-400 and its planned excellent energy resolution, the
discrimination between a monochromatic line and VIB would require up to a few
years of GC observations, whereas the discrimination between one and two lines
(in case of $\gamma\gamma+\gamma Z$ final states) could be achievable much
faster~\cite{Bergstrom:2012vd}.

A substantial substructure boost will generally be necessary even in the near future to 
measure the 130 GeV signal elsewhere in the
sky, like in dwarf galaxies~\cite{GeringerSameth:2012sr,Aleksic:2012cp}, the
EGBG~\cite{Abdo:2010dk}, galaxy clusters or Milky Way subhalos. 
On the other  hand, the recent claims for an identical signal
in galaxy clusters~\cite{Hektor:2012kc} or unidentified sources of the
2FGL~\cite{Su:2012zg} are likely to soon be confirmed or refuted in light 
of the upcoming data and improved telescope performances mentioned above.

\subsection{Fundamental reach}
\label{sec:fundamental}
Let us finally estimate what may be
called the \emph{systematic} or \emph{fundamental} reach of gamma-ray searches
for DM
 signals with current technology. As an instructive exercise, and somewhat complementary to
other works (see e.g.~Ref.~\cite{Bergstrom:2010gh}), we will here initially
assume infinite
observational time or effective area, and concentrate on the remaining systematic
limitations. These limitations come from i) a modeling of astrophysical
backgrounds, and ii) the instrument itself (see also
Ref.~\cite{Funk:2012ca} for an instructive
comparison of sensitivities of space- and ground-based telescopes).

Numerical codes like GALPROP~\cite{Strong:1998pw} do an excellent job in
modeling the galactic diffuse gamma-ray emission. However, uncertainties
related to {e.g.}~the simplified propagation set-up, the interstellar
radiation field or models for the gas distribution inhibit \textit{a
priori} predictions, and even after performing fits to the data up to
$\sim$30\% residuals at large and small scales remain~\cite{FermiLAT:2012aa,
Ackermann:2012rg}. At high latitudes, away from the galactic disk,
background variations at $\sim$1$^\circ$ scales are much less dramatic. Here, we
will adopt a (very) optimistic $1\%$ systematic background uncertainty for dwarf
galaxy and GC observations.

Relevant instrumental systematics are the spatial and spectral variations in
the effective area, incomplete rejection of cosmic rays, and uncertainties in
the energy and directional reconstruction (in case of the LAT, {e.g.}, spectral
uncertainties in the effective area range from $2\%$ to $10\%$, depending on
signature of interest~\cite{Fermi:caveatsPASS7}). We will adopt here an
optimistic reference value of $1\%$ as uncertainty for the effective area,
which still leaves room for future improvements in the instrumental design. In case
of IACTs, we furthermore adopt the aggressive scenario that \textit{all}
hadronic showers are rejected, and only the cosmic-ray electron flux remains
as an irreducible background (at energies below a few hundred GeV, this
approximation is actually already realized with current technology).
Such improvement could finally come from an improved imaging of the air
shower, and from using a large sensitive array to veto cosmic-ray induced
showers by the debris that they typically induce at relatively large angles
from the shower axis. Although we don't include this possibility in our
estimates, one has to keep in mind that even a rejection of the electron induced
background could be finally possible by detecting the Cherenkov light of
electrons before their first interaction~\cite{Consortium:2010bc}.
 
\medskip

We will in the following consider three conceptionally different targets, which are of
particular interest for current DM searches: a continuum signal from
dwarf spheroidal galaxies, a gamma-ray line signal from the GC,
and a continuum signal from the GC.


\emph{Dwarf Spheroidals.} While a combined analysis of several dwarfs could 
further improve
the results, if backgrounds and instrumental systematics are under control, 
we will here focus on a single prototypic dwarf
spheroidal galaxy. We
adopt a reference $J$-value of $J\equiv\int d\Omega\int_{\rm los}\!\!d\ell\,\rho_\chi^2=10^{19}\GeV^2 \cm^{-5}$
 inside an integration cone with radius $\theta=0.15^\circ$,
which is of the expected order for e.g.~Draco or 
Segue~I~\cite{Ackermann:2011wa, Cholis:2012am}. For
other values our limits would roughly scale like $\propto J^{-1}$ for constant $\theta$, but
 the potential cuspiness of the DM profile could be used to further strengthen
the limits by choosing  a smaller integration cone (albeit only for angular resolutions well 
below $0.1^\circ$, see e.g.~Fig.~1 in Ref.~\cite{Bringmann:2008kj}).
In case
of space-based instruments, we estimate the level of the diffuse background by
the IGRB determined by Fermi \cite{Abdo:2010nz}. For ground-based instruments, we add
on top of that the $e^\pm$-flux measured by Fermi-LAT~\cite{Ackermann:2010ij}.

At the right end of Fig.~\ref{fig:time}, with dark blue and red markers, we
show the
limits that result from the requirement that the signal flux from our 
reference dwarf is below a factor of $2\sqrt{2}\%$ 
of the background within the integration region $\theta$ at all energies
(corresponding to a $2\sigma$ error from 1\% instrumental
and background systematics).
As apparent from this plot,
especially space-based instruments still allow a substantial improvement of
the limits.
To really reach these systematic limitations,
however, one would need in case of space-based dwarf galaxy observations
an unrealistic observational time of $10^3$
($4\times10^4$, $9\times10^5$) years for DM masses of $m_\chi=10\GeV$
($100\GeV$, $1000\GeV$), which can only be overcome with a larger effective
area than the assumed $A_\text{eff}=1$\,m$^2$. For comparison, we therefore 
also show by the light red symbols the limits
that could be obtained with a hypothetical exposure about 100 times larger than 10
years of Fermi LAT observations in survey mode ($\sim 5\times10^{13}\cm^2\s$). 
Ground based telescopes -- which reach much higher event numbers
than space based instruments -- may reach the quoted `fundamental' limit
already within realistic observational times (e.g.~about 100h for
$m_\chi=1\TeV$ and $A_\text{eff}=1\km^2$).
Any improvement beyond these limits would require an efficient rejection of
cosmic-ray electrons, which is however extremely challenging (see above).

\emph{Gamma-ray lines from the Galactic Center.} 
To estimate the fundamental or systematic limit for gamma-ray line searches in case of
space-based instruments, we assume the same ROI, $J$-value and background fluxes as in
Ref.~\cite{Bergstrom:2012vd}. The ROI has a size of $\sim20^\circ$ and
includes the GC; it is optimized for a large signal-to-noise ratio. The
$J$-value derives from an Einasto profile. We adopt an energy resolution of
1\% (as expected for
e.g.~GAMMA-400). Requiring that the peak of the line signal after convolution
with the instrumental response (assumed to be Gaussian for simplicity) does not
overshoot the background flux by more than $2\sqrt{2}\%$ yields the limits
shown by the green markers at the right end of Fig.~\ref{fig:time}. These
fundamental limits are $10$--$100$ times stronger than what is currently
obtained with Fermi-LAT 
(see also Ref.~\cite{Bringmann:2011ye} for prospects of observing line or IB 
features with current and future IACTs). Again, the light green symbols show
the limits that would be obtained after a GC exposure 100 times larger than
with Fermi LAT after 10 years.

\emph{Continuum signal from Galactic Center.}
Lastly, we estimate the systematics limits for constraints on annihilation into
$\bar{b}b$ final states that can be obtained from GC observations with
space-based instruments. We adopt here the same ROI, $J$-value and background
fluxes as above. The results are shown in Fig.~\ref{fig:time} by the orange
markers. At low DM masses around $10\GeV$, they are only a factor of a few
stronger than what is already now obtained with Fermi LAT. More than an order
of magnitude could still be gained at $\TeV$ masses, which would allow to
probe DM signals down to the thermal cross-section (recall, however,  that the uncertainties of the
$J$-value are very large, as discussed above). However, to really reach
these limits on statistical grounds, one would again need a GC exposure 100
times larger than what is obtained with Fermi LAT after 10 years.

\section{Conclusions}
\label{sec:conc}

In this review, we have argued that one may consider gamma rays as the
\emph{golden channel} of indirect searches for DM in view of the
extraordinarily rich spectral and angular information they can carry. This
does not only help to discriminate signals from backgrounds but could
eventually reveal valuable details about the properties of the DM particles.
We have discussed the most important signatures in quite some detail and
provided an update on current limits, demonstrating that indirect searches
start to probe realistic cross sections and thus become competitive probes of
physics beyond the standard model.

While too early for a final judgement at the time of this writing, the line
feature at 130\,GeV that is seen in the Fermi data might turn out to be the
most promising DM \emph{signal} claimed so far. In fact, the intrinsic width of 
this feature must be smaller than  roughly 20\% (18\% at 95\%CL) -- 
which leaves  lines, VIB or box signals (Fig.~\ref{fig:spec_comp}) as 
possible channels for an explanation in terms of DM. On the other hand, it is
extremely challenging to find any explanation related to astrophysics for such a spectral 
feature; for example, even a 
very hard contribution to the gamma-ray flux, 
with a sharp break at 130\,GeV, cannot describe the data in a satisfactory way 
(Fig.~\ref{fig:powerlaw}). 
The signal morphology is perfectly consistent with annihilating DM and an 
Einasto profile for the DM density, at least for distances larger than the possible 
displacement from the GC by 1-2 degrees, but essentially rules out both cored and more contracted  
profiles (Fig.~\ref{fig:profile}); decaying DM is also in strong tension with the data. 
The data  show a weak 
hint for a second peak at 114\,GeV (Fig.~\ref{fig:gggZ} and Tab.~\ref{tab:line_limits}) 
which is exactly the combination of energies expected 
for the annihilation of 130\,GeV DM particles into $\gamma\gamma$ and $\gamma Z$ final states.
However, large annihilation rates into these channels rather generically imply
large annihilation rates rate also at tree-level, in potential conflict with continuum 
gamma-ray limits; VIB, on the other hand, does not suffer from this drawback 
(Fig.~\ref{fig:SUSY_scan}).

If confirmed by the Fermi
collaboration or other experiments, and in the absence of satisfactory
instrumental or astrophysical explanations, this signal would lead to the exciting
conclusion that the first particle beyond the standard model has been found in
space rather than at a collider. We have discussed at length how astrophysical
observations would already now help to determine detailed properties of this
new particle. 
The situation will further improve  in the relatively near future given that  prospects 
to study the 130\,GeV  feature in more detail are extremely good. However, we believe that 
even if the DM origin of the signal is eventually not confirmed,
our analysis serves to make a compelling case for the importance of focussing
on clear spectral features in future searches for DM.

During the last ten years or so, most limits on DM annihilation have improved by about 
one order of magnitude and this trend is expected to continue for the \emph{next decade} 
(Fig.~\ref{fig:time}). We have further estimated the systematics-limited (or `fundamental') reach of 
gamma-ray experiments with present technology, demonstrating that 
there is still quite some room for improvement even beyond those limits expected for
the next decade, 
especially for space-based instruments
(but also for ground-based telescopes \emph{if} the cosmic-ray electron background  can
at least partially be rejected). 
Eventually, it may thus in principle be 
possible to probe cross sections down to at least one order of magnitude below the 
thermal value for TeV-scale particles; for many models, this would correspond
to interactions too feeble  to show up in any other kind of experiment, including 
direct or collider searches. While even those limits may not be sufficient to 
completely close the window for WIMP DM, model-building would certainly need to
become increasingly sophisticated to avoid them.

Let us finally stress that in order to fully identify the properties of the DM particles,  
it will of course be indispensable to correlate a suspected DM signal  in gamma rays with 
results from indirect searches at other wavelengths and with other messengers.
The same holds for direct searches and new data from colliders, both of which
are guaranteed to deliver substantial new input in the near future -- be it in terms
of greatly improved limits or actual first hints for a signal. Chances are thus high that 
the next decade will either bring us a great deal closer to the long-sought nature of DM
or, in the most pessimistic scenario in terms of detectional prospects, force us to seriously 
question the very idea of DM being composed of WIMPs.

\bigskip
\section*{Acknowledgments}

We warmly thank Lars Bergstr\"om, Wilfried Buchm\"uller, Marco Cirelli, Jan Conrad, Michele Doro, Christian Farnier, Dmitry  Gorbunov, Michael Gustafsson, Werner Hofmann, Dieter Horns, Michael Kuhlen, Julien Lavalle, Manfred Lindner, Pat Scott, Thomas Schwetz,
Pasquale Serpico, Joe Silk, Peter Tinyakov, Nicola Tomassetti, Nikolay Topchiev and Gabrijela
Zaharijas for valuable discussions and
feedback on the manu\-script.  T.B.~gratefully acknowledges support from the
German Research Foundation (DFG) through the Emmy Noether  grant BR 3954/1-1.
T.B.~also would like to thank the Institut d'Astrophysique de Paris, and especially Joe Silk, for hospitality during the initial stages of this work.  C.W.~acknowledges partial support from the European 1231 Union FP7 ITN
INVISIBLES (Marie Curie Actions, PITN-GA-2011-289442). 

\newpage

\bibliographystyle{JHEP_mod}

\bibliography{review}

\end{document}